\documentclass[10pt,journal]{IEEEtran}

\ifCLASSOPTIONcompsoc
  \usepackage[nocompress]{cite}
\else
  \usepackage{cite}
\fi

\usepackage{amsmath,amssymb,amsfonts}
\usepackage{algorithmic}
\usepackage{graphicx}
\usepackage{textcomp}
\usepackage{xcolor}
\usepackage{bm}
\usepackage{booktabs}
\usepackage{array}
\usepackage{diagbox} 
\usepackage{caption}
\usepackage{colortbl}
\usepackage{tabu}
\usepackage{subfigure}
\usepackage{soul}
\usepackage[numbers,sort&compress]{natbib}
\usepackage{lipsum}
\usepackage{mathtools}
\usepackage{cuted}

\ifCLASSINFOpdf
\else
\fi

\begin{document}
%
\title{Non-Uniform Time-Step Deep Q-Network for Carrier-Sense Multiple Access in Heterogeneous Wireless Networks}
\author{Yiding Yu,~\IEEEmembership{Student Member,~IEEE,}
        Soung Chang Liew,~\IEEEmembership{Fellow,~IEEE,} \\
        and Taotao Wang,~\IEEEmembership{Member,~IEEE}
\IEEEcompsocitemizethanks{
\IEEEcompsocthanksitem Y. Yu and S.C. Liew are with the Department of Information Engineering, The Chinese University of Hong Kong, Shatin, Hong Kong Special Administrative Region, China. Email: \{yy016, soung\}@ie.cuhk.edu.hk 
\IEEEcompsocthanksitem T. Wang was with the Department of Information Engineering, The Chinese University of Hong Kong, Shatin, Hong Kong Special Administrative Region, China. He is now with the College
of Information Engineering, Shenzhen University, Shenzhen 518061, China. Email: ttwang@ie.cuhk.edu.hk
}
}

\IEEEtitleabstractindextext{%
\begin{abstract} 
This paper investigates a new class of carrier-sense multiple access (CSMA) protocols that employ deep reinforcement learning (DRL) techniques, referred to as carrier-sense deep-reinforcement learning multiple access (CS-DLMA). The goal of CS-DLMA is to enable efficient and equitable spectrum sharing among a group of co-located heterogeneous wireless networks. Existing CSMA protocols, such as the medium access control (MAC) of WiFi, are designed for a homogeneous network in which all nodes adopt the same protocol. Such protocols suffer from severe performance degradation in a heterogeneous environment where there are nodes adopting other MAC protocols. CS-DLMA aims to circumvent this problem by making use of DRL. In particular, this paper adopts  $\alpha$-fairness as the general objective of CS-DLMA. With  $\alpha$-fairness,  CS-DLMA can achieve a range of different objectives (e.g., maximizing sum throughput, achieving proportional fairness, or achieving max-min fairness) when coexisting with other MACs by changing the value of  $\alpha$. A salient feature of CS-DLMA is that it can achieve these objectives without knowing the coexisting MACs through a learning process based on DRL. The underpinning DRL technique in CS-DLMA is deep Q-network (DQN). However, the conventional DQN algorithms are not suitable for CS-DLMA due to their uniform time-step assumption. In CSMA protocols,  time steps are non-uniform in that the time duration required for carrier sensing is smaller than the duration of data transmission. This paper introduces a non-uniform time-step formulation of DQN to address this issue. Our simulation results show that CS-DLMA can achieve the general   $\alpha$-fairness objective when coexisting with TDMA, ALOHA, and WiFi protocols by adjusting its own transmission strategy. Interestingly, we also find that CS-DLMA is more Pareto efficient than other CSMA protocols, e.g., p-persistent CSMA, when coexisting with WiFi. Although this paper focuses on the use of our non-uniform time-step DQN formulation in wireless networking, we believe this new DQN formulation can also find use in other domains.
\end{abstract}
}

\maketitle

\IEEEdisplaynontitleabstractindextext
\IEEEpeerreviewmaketitle

\section{Introduction}\label{sec:introduction}
This paper considers the problem of efficient and equitable spectrum sharing among a group of co-located heterogeneous wireless networks. These networks may adopt different medium access control (MAC) protocols and they do not know the MAC protocols of other networks. This scenario is envisioned by DARPA Spectrum Collaboration Challenge (SC2) competition as a future spectrum sharing paradigm \cite{DARPAwebsite, tilghman2019will}. In this futuristic scenario, unlike in the conventional cognitive radio,  all users/networks are on equal footing in that they are not divided into primaries and secondaries. When sharing the spectrum  in an efficient and equitable manner, each network must respect spectrum usages by other networks in that it must not hog the spectrum to the detriment of other networks. A major \textbf{challenge} for one particular network is \textit{how to coexist with other networks without knowing the MACs of other networks while achieving efficient and equitable spectrum usage among all networks}. 
\begin{figure}[!t]
	\centering
	\includegraphics[scale=0.7]{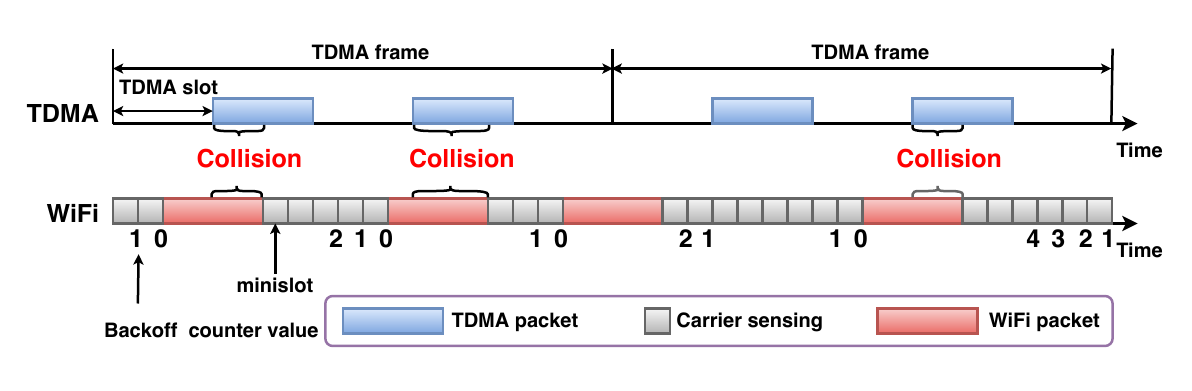}
	\caption{Inharmonious coexistence of a TDMA node and a WiFi node. For simplicity, this example assumes each TDMA slot/packet and each WiFi packets last four minislots, where a minislot is the slot used for carrier sensing by WiFi. TDMA transmits packets in specific time slots within a TDMA frame repeatedly, from frame to frame, regardless of the MAC of WiFi. When WiFi senses the carrier to be idle and transmits in the subsequent four minislots, its transmission may collide with a TDMA packet that follows shortly.}
	\label{fig:illustration}
	\vspace{-0.20in}
\end{figure}

Widely used wireless MAC protocols today are often designed for homogeneous networks in which all nodes use the same MAC. A case in point is WiFi, which adopts a particular form of  carrier-sense multiple-access (CSMA) with collision avoidance \cite{Tanenbaum:2010:CN:1942194}. The carrier sensing and binary exponential backoff mechanisms of WiFi \cite{bianchi2000performance, liew2010back} work well only if all nodes in the network adopt the same mechanisms. They do not work well in heterogeneous networks. To illustrate, consider the coexistence of a WiFi node and a node operating the time division multiple access (TDMA) protocol. The TDMA node transmits in specific time slots in a frame consisting of multiple time slots, in a repetitive manner from frame to frame, as illustrated in Fig. \ref{fig:illustration}. In particular, the TDMA channel access pattern is oblivious of the MAC of WiFi; similarly, the MAC of WiFi is oblivious of the TDMA channel access pattern. As shown in Fig. \ref{fig:illustration}, the WiFi node may sense the channel to be idle and decide to transmit a packet, only to have the TDMA node transmit a packet shortly thereafter to result in a collision. This leads to inefficiency of the spectrum usage in the heterogeneous network setting. 

A goal of this paper is to circumvent this problem with a new class of CSMA protocols based on deep reinforcement learning (DRL) \cite{sutton2018reinforcement}. DRL is a machine learning technique that combines deep learning and reinforcement learning. DRL has had success in solving a wide range of complex decision-making tasks, including video game playing, robotic control, smart grid management, and wireless communication \cite{li2017deep, DQNpaper, silver2016mastering, gu2017deep, ruelens2016residential, luong2019applications, sun2019application}. A salient feature of our DRL-based MAC protocol is that \textit{it does not need to know the operation mechanism of the coexisting MACs---it learns to coexist harmoniously with other MACs by trial-and-error}. Throughout this paper, the DRL-based MAC protocol is referred to as Carrier-Sense Deep-reinforcement Learning Multiple Access (CS-DLMA). The nodes operating CS-DLMA are referred to as CS-DLMA nodes and the corresponding radio network is referred to as CS-DLMA network.

In general, CS-DLMA can have different objectives when coexisting with other MACs, e.g., maximize sum throughput, achieve proportional fairness or achieve max-min fairness \cite{mo2000fair}. For generality, this paper adopts  $\alpha$-fairness \cite{mo2000fair} as the objective of CS-DLMA. With   $\alpha$-fairness, CS-DLMA can achieve a range of different objectives by changing the value of  $\alpha$, including the aforementioned objectives. We show that CS-DLMA can achieve near-optimal results with respect to different  $\alpha$ values when coexisting with other MAC protocols, such as TDMA,  ALOHA, and WiFi. Moreover, we demonstrate that CS-DLMA is more Pareto efficient \cite{myerson2013game} than p-persistent CSMA \cite{Tanenbaum:2010:CN:1942194} when coexisting with WiFi.

The underpinning DRL technique in CS-DLMA is deep Q-network (DQN) \cite{DQNpaper}, developed by DeepMind to achieve superhuman level performance in playing Atari games. However, the original DQN in \cite{DQNpaper} is not directly applicable for our propose for two reasons:

\begin{enumerate}
	\item The original DQN only aims to maximize the cumulative discounted ``rewards''---i.e., the objective or ``return'' to be optimized is a weighted linear sum of the rewards in consecutive time steps \cite{sutton2018reinforcement}---and this does not fit in with the  $\alpha$-fairness objective, which in general is a nonlinear combination of utility functions.
	\item The original DQN is built on a discrete-time framework wherein an underlying assumption is that the time steps are of uniform duration. Specifically, this implicit assumption is made in the way that it discounts the rewards from time step to time step in a uniform manner. For CSMA protocols, the time slots are non-uniform in nature in that the minislots used for carrier sensing are of smaller duration than time slots used for data transmission. 
\end{enumerate}

Our previous work \cite{yu2019deep} put forth a multi-dimensional DQN algorithm to solve issue 1). This paper introduces a non-uniform time-step formulation in DQN to address issue 2). The key idea in non-uniform time-step DQN is that we need to discount “reward” according to the duration of each time step.\footnote{We remark that our non-uniform time-step DQN formulation is also potentially applicable to other decision-making problems. For example, in the problem of Treasury bond investment, the maturity dates and the interest rates of different bonds may be different. If the investment strategy is to decide the maturity date of the bond to be purchased based on certain observed ``environmental states'', then the time duration between successive decision-making/investment epochs may vary according to the maturity dates of the bonds. To discount properly, the DRL agent needs to take into account the different time durations.}

\subsection{Contributions}
We summarize our contributions in this paper as follows:
\begin{itemize}
	\item We develop a new class of CSMA protocol based on DRL, referred to as CS-DLMA, for spectrum sharing in heterogeneous wireless networks. A salient feature of CS-DLMA is that it not only optimizes its own throughput but also the throughputs of other coexisting networks according to the general  $\alpha$-fairness objective.  Importantly, CS-DLMA achieves this without knowing the MAC protocols of other networks. 
	\item We demonstrate that CS-DLMA can achieve the general  $\alpha$-fairness objective when coexisting with the TDMA, ALOHA, and WiFi protocols by adjusting its own transmission strategies. Interestingly, we find that CS-DLMA is more Pareto efficient than other CSMA protocols, e.g., p-persistent CSMA, when coexisting with WiFi. 
	\item We put forth a non-uniform time-step multi-dimensional DQN algorithm to enable CS-DLMA to achieve the above performance. Although we only focus on the use of the modified DQN algorithm for wireless networking, we believe it can also find use in other domains with similar non-uniform time-step and multi-dimensional characteristics. 
\end{itemize}
\subsection{Related Work}
Since this paper focuses on MAC designs based on DRL techniques, we limit our review of related work in the same area only. A substantial body of past related work on DRL based MAC, for example, \cite{naparstek2018deep, wang2018deep, chang2018distributive, zhong2018actor, xu2018deep}, did not consider MAC with carrier sensing. In the set-up of the past work, the time steps in decision making are with the same duration. For MAC with carrier sensing, the carrier sensing time and the packet transmission time are of different durations, and the conventional DRL algorithms with the implicit assumption of uniform time steps are not suitable anymore. This is the reason why in the current paper of ours, we need to modify the conventional DRL techniques so that they can be used for the design of DRL MAC with carrier sensing and for the coexistence of the DRL MAC with other MACs with the carrier sensing capability. 

The authors in \cite{challita2018proactive} and \cite{tan2019deep} investigated the coexistence of the DRL based LTE network with WiFi, which has the carrier sensing capability.  However, the LTE MACs in \cite{challita2018proactive} and  \cite{tan2019deep} exercise coarse-grained control in that the decisions are not made on a packet-by-packet basis. Specifically, the LTE MAC does not decide whether to transmit on a packet-by-packet basis, but rather decides a stretch of time for LTE transmissions and a stretch of time for WiFi transmissions. During the respect stretches of time, LTE/WiFi get to keep transmitting packets without interruption from the other network. By contrast, the MAC of our design in this paper exercises fine-grained control in the MAC decides whether to transmit the next packet based on carrier sensing the environment as well as the past history of the environmental state. 

In the following, we elaborate other fine differences between our work and \cite{naparstek2018deep, wang2018deep, chang2018distributive, zhong2018actor, xu2018deep, challita2018proactive, tan2019deep}. The DRL MAC proposed in \cite{naparstek2018deep} is targeted for homogeneous wireless networks. Specifically, in \cite{naparstek2018deep}, multiple nodes access multiple time-invariant orthogonal channels using the same DRL MAC. By contrast, we focus on heterogeneous networks in which our CS-DLMA protocol must learn to coexist with other MAC protocols. The MACs in \cite{wang2018deep, chang2018distributive, zhong2018actor, xu2018deep} are also concerned with multiple-channel access. Unlike \cite{naparstek2018deep}, the channels in \cite{wang2018deep, chang2018distributive, zhong2018actor, xu2018deep} are time varying and the channels may be occupied by some ``primary'' or ``legacy'' nodes. The DRL nodes aim to maximize their own throughputs by learning the channel characteristics and the transmission patterns of the ``primary'' or ``legacy'' nodes. By contrast, the CS-DLMA nodes in our work aim to achieve a global  $\alpha$-fairness objective, which includes achieving maximum sum throughput, proportional fairness, and max-min fairness as subcases.

Both the MAC schemes in \cite{challita2018proactive} and \cite{tan2019deep} are model-aware in that the LTE base stations know that the coexisting network is WiFi. Therefore, the approaches in \cite{challita2018proactive} and \cite{tan2019deep} are not generalizable to situations where the LTE stations coexist with other networks. For example, suppose that instead of WiFi, the other network is ALOHA. Given that ALOHA does not perform carrier sensing, an ALOHA node may still transmit while an LTE node transmits during the stretch of time allocated to LTE, leading to collisions. By contrast, our CS-DLMA protocol is model-free in that it does not presume knowledge of coexisting networks---our CS-DLMA protocol can coexist with any MAC protocol by nature. 

In our previous work \cite{yu2019deep}, we developed deep reinforcement learning multiple access (DLMA) protocols for heterogeneous networking without carrier sensing. Furthermore, we assumed that nodes of different MACs use the same packet length. This assumption limits the application of DLMA in more general heterogeneous settings in which nodes of different MACs may adopt different packet lengths. Our early work \cite{yu2018carrier} incorporated carrier sensing into DLMA. However, for simplicity, \cite{yu2018carrier} assumed the durations of carrier sensing and packet transmissions of DRL nodes are the same. Our current paper removes this impractical assumption. As a result, the DRL time steps are of non-uniform durations now. We put forth a non-uniform time-step formulation of the DQN algorithm to address the issue.
\section{Reinforcement Learning Preliminaries}
This section overviews the reinforcement learning (RL) techniques \cite{sutton2018reinforcement}. In the RL framework, a decision-making agent interacts with an environment in discrete time steps. At time step  $ t $, the agent observes the environment state   $ s_t $ and performs an action  $ a_t $  chosen from an action set   according to a policy   $ \pi $. The policy  $ \pi $  is a mapping from states to actions. Following the action  $ a_t $, the agent receives a reward  $ r_{t+1} $ and the environment transits to state $ s_{t+1} $   at time step  $ t+1 $. There are different techniques for reinforcement learning. This paper adapts and extends the Q-learning technique \cite{watkins1992q} for our particular application. 
\subsection{Q-learning}
Given a series of rewards,  $r_{t + 1}, r_{t + 2}, \cdots $, resulting from state-action pairs  $\left( s_t, a_t\right) , \left( s_{t+1}, a_{t+1}\right) , \cdots$, for Q-learning, the cumulative discounted return going forward pinned at time step $ t $  is given by  $ {R_t} \buildrel \Delta \over = \sum\nolimits_{k = 0}^\infty  {{\gamma ^k}{r_{t + k + 1}}} $, where  $\gamma  \in \left( {0,1} \right]$  is a discount factor. Because of the randomness in the state transitions,  ${R_t}$  is a random variable in general. Q-learning captures the expected cumulative discounted reward of a state-action pair $ \left( s, a\right) $  of a policy $ \pi $  by a Q action-value function:   ${Q^\pi }\left( {s,a} \right) \buildrel \Delta \over = \mathbf{E}\left[ {{R_t}|{s_t} = s,{a_t} = a,\pi } \right]$. The Q function of an optimal policy among all policies is  ${Q^*}\left( {s,a} \right) \buildrel \Delta \over = {\max _\pi }{Q^\pi }\left( {s,a} \right)$.  

In Q-learning, the goal of the agent is to learn the optimal policy in an online manner by observing the rewards while taking action in successive time steps. In particular, the agent maintains the Q function,  $Q\left( s, a\right)$, for any state-action pair   $ \left( s, a\right) $,  in a tabular form. At time step   $ t $,  given state  $ s_t $, the agent selects an action ${a_t} = \arg {\max _a}Q({s_t},a)$  based on its current estimated Q table. This will cause the system to return a reward  $ r_{t+1} $ and move to state  $ s_{t+1} $. The experience at time step   $ t $ is captured by the quadruplet  ${e_t} = ({s_t},{a_t},{r_{t + 1}},{s_{t + 1}})$. At the end of time step  $ t $, experience $ e_t $  is used to update  $Q({s_t},{a_t})$ for entry $({s_t},{a_t})$  as follows: 	
\begin{align}\label{eqn1}
{Q_{new}}&\left( {{s_t},{a_t}} \right) \leftarrow {Q_{old}}\left( {{s_t},{a_t}} \right) +\nonumber \\
&\beta \left[ {{r_{t + 1}} + \gamma \mathop {\max }\limits_{a'} {Q_{old}}\left( {{s_{t + 1}},a'} \right) - {Q_{old}}\left( {{s_t},{a_t}} \right)} \right],
\end{align}	
where $\beta  \in (0,1]$ is referred to as the learning rate. 

In Q-learning, the so-called  $ \varepsilon $-greedy algorithm is often adopted for action selection. For the  $ \varepsilon $-greedy algorithm, the action  ${a_t} = \arg {\max _a}Q({s_t},a)$  is  chosen with probability   $ 1-\varepsilon $, and a random action is chosen uniformly among all possible actions with probability  $ \varepsilon $.  The random action is incorporated to avoid the algorithm from zooming into a local optimal policy and to allow the agent to explore a wider spectrum of different actions in search of the optimal policy, particularly at the early stage of the learning process. 

Q-learning is a \textit{model-free} learning framework in that it tries to learn the optimal policy without having a model that describes the operating behavior of the environment beyond what can be observed through the experiences. In particular, it does not have knowledge of the transition probability  $P[{s_{t + 1}},{r_{t + 1}}|{s_t},{a_t}]$. 

\subsection{Deep Q-Network}\label{Sec:one-step DQN}
It has been shown that in a stationary environment that can be fully captured by a Markov decision process, the Q-values will converge to the optimal ${Q^*}\left( {s,a} \right)$   if the learning rate decays appropriately and each action in the state-action pair  $\left( {s,a} \right)$  is executed an infinite number of times in the process \cite{watkins1992q}. For many real-world problems, the state-action space for $\left( {s,a} \right)$  can be huge that the tabular update method, which updates only one entry in  $Q\left( {s,a} \right)$  in each time step, can take an excessive amount of time for   $Q\left( {s,a} \right)$ to converge to  ${Q^*}\left( {s,a} \right)$. If the environment changes in the meantime (e.g., $P[{s_{t + 1}},{r_{t + 1}}|{s_t},{a_t}]$ changes), convergence can never be attained. To allow fast convergence, function approximation methods are often used to approximate the Q-values \cite{sutton2018reinforcement}.

The seminal work \cite{DQNpaper} put forth an algorithm referred to as the deep Q-network (DQN), wherein a deep neural network model is used to approximate the action-value function  $ Q $. To avoid confusion between the algorithm DQN from the neural network used in the algorithm, in this paper we refer to the neural network as the Q neural network (QNN). For the same algorithm, different possible QNNs could be used. 

The input to a QNN is a state  $ s $, and the outputs are the approximated Q-values for different actions,  $ \left\{ {Q\left( {s,a;{\bm{\theta }}} \right)|a \in \mathbb{A}} \right\} $, where ${\bm{\theta }}$   is a parameter vector consisting of the weights of the edges in the neural network and $\mathbb{A}$ is the set of possible actions.  At the end of time step  $ t $, for action execution, the  $ \varepsilon $-greedy algorithm, wherein  ${a_t} = \arg {\max _a}Q({s_t},a)$ is replaced by ${a_t} = \arg {\max _a}Q({s_t},a;{\bm{\theta }})$, is adopted.  

For training of the QNN, the parameters  of the QNN,   ${\bm{\theta }}$, are updated by minimizing the following loss function:
\begin{align}\label{eqn2}
L\left( {\bm{\theta }} \right) = \frac{1}{{{N_E}}}\sum\limits_{{e_\tau } \in E} \Bigl( {r_{\tau  + 1}} & + \gamma \mathop {\max }\limits_{a'} Q\left( {{s_{\tau  + 1}},a';{{\bm{\theta }}^ - }} \right)\nonumber \\ & - Q\left( {{s_\tau },{a_\tau };{\bm{\theta }}} \right) \Bigr)^2.
\end{align}
In \eqref{eqn2}, two important learning techniques in DQN are embedded  to stabilize the algorithm \cite{DQNpaper}. The first is ``experience replay'' \cite{lin1992self, DQNpaper}. Instead of training QNN with a single experience at each time step, multiple experiences are pooled together for batch training. Specifically, a FIFO experience buffer is used to store a fixed number of experiences gathered from different time steps. For a round of training, a minibatch $E$  consisting of ${N_E}$  random experiences are taken from the experience buffer in the computation of \eqref{eqn2}, wherein the time index  $\tau $ denotes the time step at which that experience tuple $ {e_\tau } $  was collected.  The second technique is the use of a separate ``target neural network'' in the computation of ${r_{\tau  + 1}} + \gamma {\max _{a'}}Q\left( {{s_{\tau  + 1}},a';{{\bm{\theta }}^ - }} \right)$ in \eqref{eqn2}. In particular, the target neural network's parameter vector is ${{\bm{\theta }}^ - }$  rather than ${\bm{\theta }}$  in the QNN being trained. This separate target neural network is named target QNN and is a copy of a previously used QNN. The parameter ${{\bm{\theta }}^ - }$  of target QNN is updated to the latest  ${\bm{\theta }}$ of QNN once in a while. 

\section{System Model and Objective}
This section first introduces the system model used in this paper. After that, we give the overall system objective. 
\subsection{System Model}\label{sec:system_model}
\begin{table*}[t]
	\renewcommand\arraystretch{1.3}
	\caption{MAC mechanisms of different nodes.}
	\label{table:MAC}
	\vspace{-0.05in}
	\flushleft
	\scalebox{1}{
		\begin{tabular}{ |p{1.7cm}|p{15.4cm} |}
			\hline
			\textbf{Node Type} &\textbf{ Description} \\ \hline 
			{\footnotesize \textbf{TDMA}} &  \textit{{\footnotesize TDMA transmits in $ X $ specific TDMA slots within a TDMA frame of $ Y $  TDMA slots in a repetitive manner from frame to frame.}} \\ \hline 
			{\footnotesize \textbf{ALOHA}}  & \textit{{\footnotesize ALOHA transmits packets with a fixed probability  $q$ in each ALOHA slot in an i.i.d. manner from ALOHA slot to ALOHA slot. }} \\ 
			\hline
			{\footnotesize \textbf{WiFi}} & \textit{{\footnotesize WiFi employs a CSMA/CA protocol with the binary exponential backoff mechanism \cite{Tanenbaum:2010:CN:1942194}.  Before transmitting a packet, WiFi generates a random backoff counter value  $w \in \left[ {0,W - 1} \right]$  and performs carrier sensing on a minislot basis. For each minislot the channel is sensed idle, the backoff counter is decreased by one. The countdown of the counter is frozen if the channel is sensed busy. When the counter value reaches zero, the WiFi node transmits. The window size $W$ is doubled each time its transmission incurs a collision up to a ceiling window size of  ${2^m}W$, where  $m$ is the ``maximum backoff stage''. Upon a successful transmission, the window size reverts to the initial window size  $W$.
			}}\\
			\hline
			{\footnotesize \textbf{CS-DLMA}} & \textit{{\footnotesize The CS-DLMA node uses our CS-DLMA protocol to decide whether to perform carrier sensing or to transmit packets. If CS-DLMA decides to perform carrier sensing, it will check if the channel is busy or idle in the next minislot. If CS-DLMA decides to transmit a packet, it will also decide the packet length  $R_C$ beforehand and then transmit the packet in the next $R_C$  minislots. CS-DLMA repeats this process after carrier sensing or a packet transmission. }}    \\
			\hline
	\end{tabular}}
\end{table*}

\begin{figure}[!t]
	\centering
	\includegraphics[scale=0.5]{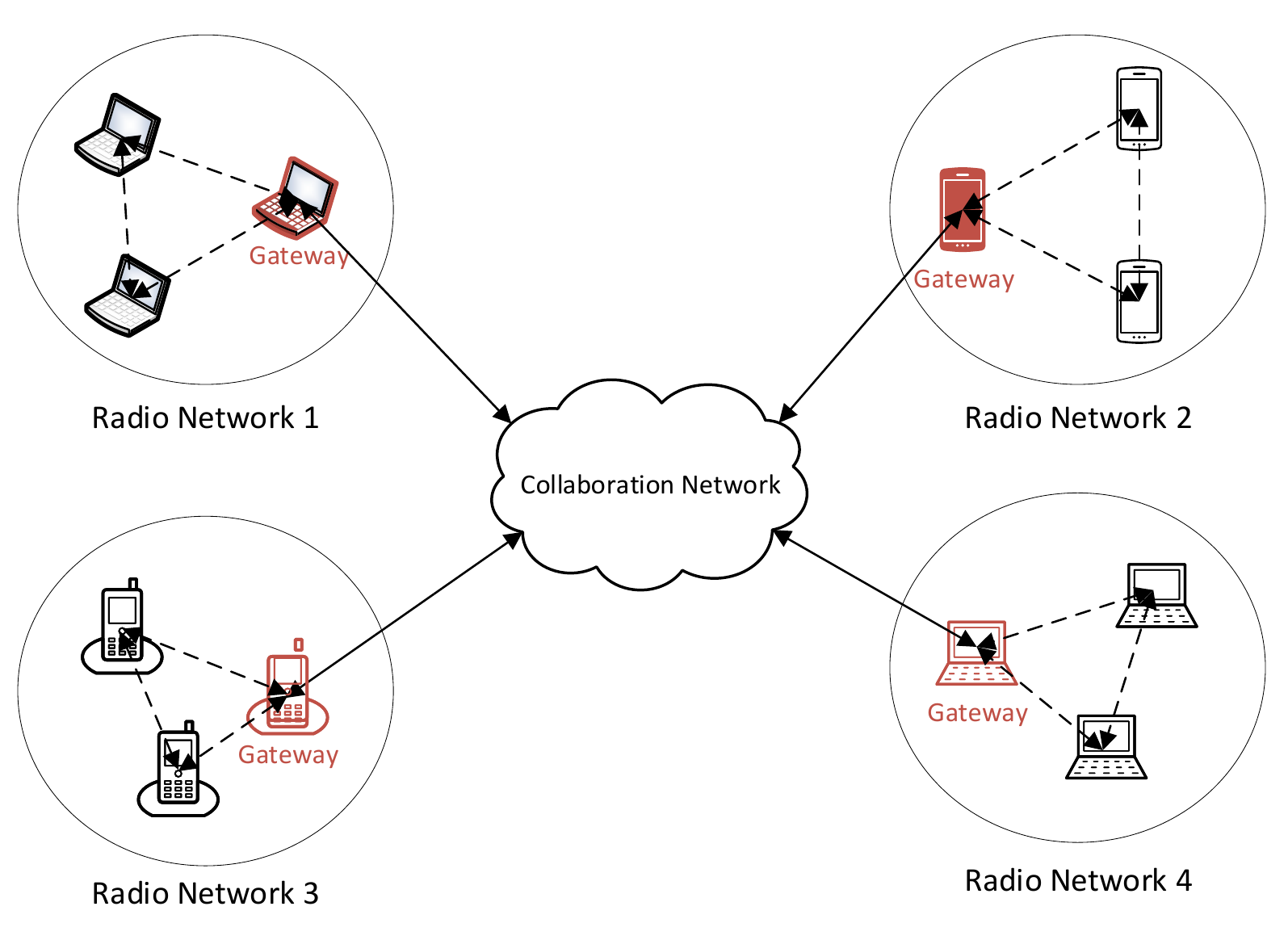}
	\caption{Network model of DARPA Spectrum Collaboration Challenge.}
	\label{fig:system_model}
\end{figure}

The system model considered in this paper is inspired by the network model of DARPA Spectrum Collaboration Challenge (SC2) \cite{DARPAwebsite, tilghman2019will}.  As illustrated in Fig. \ref{fig:system_model}, the model of DARPA SC2 is composed of a collaboration network and multiple radio networks. All the radio networks share a common wireless medium. In DARPA SC2, the collaboration network is a separate control network from the wireless data network. The collaboration network allows different radio networks to communicate collaborative information at the high level (e.g., the frequency spectrum used by a radio network, the throughput and the quality of service observed in a radio network, etc.). Each radio network, however, does not tell the other networks its MAC protocol. 

On the wireless data channel, the nodes in each radio network can transmit data packets to each other, whereas the nodes belonging to different radio networks do not exchange data packets. A packet is deemed to be successfully transmitted if there are no concurrent transmissions by other nodes. Otherwise, the packet is deemed to be lost due to a collision.

Each radio network operates a MAC protocol that determines the transmission strategy of its nodes.  The coexisting radio networks are \textit{heterogeneous} in that they may adopt different MAC protocols. Importantly, each radio network does not know the MAC protocols of other radio networks. The goal, in DARPA's vision, is to optimize the aggregate wireless spectrum usage across all radio networks \cite{DARPAwebsite, tilghman2019will}.

An important feature of DARPA SC2's model is that in each radio network, a node is designated as a gateway for collaborative information exchange with gateways of other radio networks through the collaboration network. In this work, as will be elaborated later, we assume the collaborative information includes transmission results of networks, such as successes/failures of packet transmissions and packet durations. The gateway of a radio network may in turn share the transmission results of other radio networks with its own nodes. Using collaborative information, a radio network can then adjust its transmission strategy through an adaptive MAC protocol to achieve a certain global objective to share the wireless spectrum with other networks in an equitable and optimal manner.  For example, if the objective is to achieve proportional fairness, the adaptive MAC protocol will aim to maximize the sum log throughputs of all networks \cite{yu2019deep}.  

This paper focuses on the design of the MAC protocol of a particular radio network. The goal is to be able to \textit{achieve a general global objective for wireless-spectrum sharing without knowing the MAC protocols of other radio networks}. 

We assume the nodes in our radio network have \textit{carrier-sensing} (CS) capability, and our MAC protocol exploits deep reinforcement learning techniques to learn a transmission strategy that can achieve the global objective. We refer to our MAC protocol as Carrier-Sense Deep-reinforcement Learning Multiple Access (CS-DLMA). Our network and nodes are referred to as CS-DLMA network and CS-DLMA nodes.

Carrier sensing allows radio nodes to listen to the wireless channel before transmitting their data packets  so as to avoid collisions \cite{Tanenbaum:2010:CN:1942194}.  The carrier sensing operation typically takes up some time that includes the signal processing and circuit delay within a node, as well as the largest possible signal propagation over the air between nodes. To be effective, carrier sensing time must be small relative to the data packet duration. In this paper, we refer to the time required for carrier sensing as a ``minislot''. 

We consider three types of MAC protocols used by other radio networks: (i) TDMA, (ii) ALOHA, (iii) WiFi (more exactly, a simplified WiFi-like CSMA protocol) \cite{Tanenbaum:2010:CN:1942194}. Among these protocols, WiFi has the capability of carrier sensing, while TDMA and ALOHA do not. For simplicity, we assume the minislots used by CS-DLMA and WiFi nodes for carrier sensing are of the same duration. In addition, we assume that packet durations of different networks are integer multiples of minislots (note that packet duration here refers to the time needed to transmit MAC-layer packet header plus the data). Specifically,   ${R_C}$/${R_W}$/${R_T}$/${R_A}$   represent the packet durations of CS-DLMA/WiFi/TDMA/ALOHA. The durations of the packet headers of all networks are assumed to be the same. In particular, the packet-header duration $ H $  is a fraction of minislot ($H = 0.5$  is used in our later evaluations).  

We allow the packet duration of CS-DLMA,  ${R_C}$, to vary in time as part of its adaptive strategy. Variable ${R_C}$  gives added flexibility in CS-DLMA. For example,  if the channel is deemed to be not likely used by others for a long duration of time,  CS-DLMA can transmit a large packet with a large  ${R_C}$ to reduce the packet-header overhead and carrier-sensing overhead; a small  ${R_C}$, on the other hand,  allows CS-DLMA to squeeze in a small packet transmission  ${R_C}$ in between transmissions by others.


We summarize the MAC protocols of different networks in Table \ref{table:MAC}. We assume slotted operations of TDMA/ALOHA. A TDMA/ALOHA network can only initiate a transmission at the beginning of a TDMA/ALOHA slot, and the transmission ends at the end of a TDMA/ALOHA slot. Thus, each TDMA/ALOHA slot lasts a TDMA/ALOHA packet duration (e.g., a TDMA slot is of $R_T$ minislots in duration) in Table \ref{table:MAC}.

\subsection{$\alpha$-Fairness Objective}
We adopt the general  $\alpha$-fairness objective as the performance metric of this paper \cite{mo2000fair}.  In particular, we assume that there are altogether $N$  nodes in the overall heterogeneous wireless networks. For a particular node  $i$, its local utility function is given by
\begin{align}\label{eqn3}
{f_\alpha }\left( {{x_i}} \right) = \left\{ {\begin{array}{*{20}{c}}
	{\log \left( {{x_i}} \right),}&{{\rm{if  }} \ \alpha  = 1},\\
	{{{\left( {1 - \alpha } \right)}^{ - 1}}  {{\left( {{x_i}} \right)}^{1 - \alpha }},}&{{\rm{if   }} \ \alpha  \in \left[ {0,\left. 1 \right)} \right. \cup \left( {1,\infty } \right)},
	\end{array}} \right.
\end{align}
where  $\alpha$ is used to specify a range of fairness criteria and $x_i$  is the throughput of node  $i$. 

The objective of the overall system is to maximize the sum of all the local utility functions: 
\begin{align}\label{eqn4}
\begin{array}{l}
{\rm{maximize}} \quad F\left( {{x_1},{x_2}, \ldots ,{x_N}} \right) = \sum\limits_{i = 1}^N {{f_\alpha }\left( {{x_i}} \right)}, \\
{\rm{subject \ to     }} \quad \sum\limits_{i = 1}^N {{x_i}}  \le 1    \ {\rm{and}} \    {x_i} \ge 0, \ \forall i.
\end{array}
\end{align}
In \eqref{eqn4}, when  $\alpha=0$, the objective is to maximize the sum throughput; when  $\alpha=1$, the objective is to achieve proportional fairness; when  $\alpha  \to \infty $, the objective is to achieve max-min fairness \cite{mo2000fair}.

\section{CS-DLMA Framework}\label{sec:one-node}
This section first transforms the multiple access problem faced by our CS-DLMA network to an RL problem by defining \textbf{action}, \textbf{state} and \textbf{reward}---three key components in RL. We then modify the original DQN algorithm and put forth a non-uniform time-step multi-dimensional DQN algorithm that realizes CS-DLMA---the original DQN deals with uniform time-step problems. After that, we discuss the implementations of CS-DLMA. For simple exposition, we assume there is only one CS-DLMA node in this section. We will extend the framework to the case with multiple CS-DLMA nodes in Section \ref{sec:multi-node}. 

\subsection{Action, State, and Reward}\label{sec:action_state_reward}
\subsubsection{Action} As described in Table \ref{table:MAC}, the possible decisions of a CS-DLMA node include 1) performing carrier sensing and 2) transmitting a packet with a length of  $R_C$. We denote the action of a CS-DLMA node at time step  $t$ by  ${a_t} \in \left\{ {0,1,2, \ldots ,{R_{C\max }}} \right\}$, where ${R_{C\max }}$  is the maximum packet length of CS-DLMA (a ``time step'' here corresponds to a decision epoch of CS-DLMA and the duration of each time step can be either one minislot or multiple minislots). If  ${a_t} = 0$, the CS-DLMA node will not transmit and will only perform carrier sensing in the next minislot. The carrier sensing results in an observation ${z_t} = $  \textit{BUSY} or \textit{IDLE}, indicating whether the channel was occupied or not occupied by other nodes in that minislot. If  ${a_t} = {R_C} \in \left\{ {1,2, \ldots ,{R_{C\max }}} \right\}$, the CS-DLMA node will transmit a packet with a length of ${R_C}$  in the next ${R_C}$  minislots. At the end of the transmission, ${z_t} = $ \textit{SUCCESSFUL} or \textit{COLLIDED} will be observed, indicating whether the packet was successfully received or not. As long as another node transmits in at least one of the  ${R_C}$ minislots,  ${z_t} = $ \textit{COLLIDED} would be observed. 

\subsubsection{State} We first define the channel state of CS-DLMA at time step  $t+1$ as the action-observation pair  ${c_{t + 1}} \buildrel \Delta \over = \left( {{a_t},{z_t}} \right)$. We then define the state of CS-DLMA at time step $t+1$   as  ${s_{t + 1}} \buildrel \Delta \over = \left[ {{c_{t - M + 2}}, \cdots ,{c_t},{c_{t + 1}}} \right]$, i.e., the state is the combination of the past $M$   channel states. The state history length $M$   is the number of past time steps to be tracked by CS-DLMA. 

\subsubsection{Reward} In the conventional RL framework, the reward is a scalar and the RL agent learns to maximize the cumulative discounted reward \cite{sutton2018reinforcement}, which is a weighted linear sum of rewards in the time steps going forward.  The goal of CS-DLMA, however, is to achieve  $\alpha$-fairness among all the nodes, which in general is a nonlinear function of the individual cumulative discounted rewards (i.e. individual throughputs) of the nodes. We use a reward vector to keep track of the individual rewards in each time step, from which we can obtain the individual cumulative discounted rewards for the computation of the  $\alpha$-fairness objective function.  Specifically, after taking action  ${a_t}$, a reward vector ${{\bm{r}}_{{\bm{t + 1}}}} = \left[ {r_{t + 1}^{\left( 0 \right)},r_{t + 1}^{\left( 1 \right)},r_{t + 1}^{\left( 2 \right)}, \ldots ,r_{t + 1}^{\left( L \right)}} \right]$  is obtained from the environment at the end of time step  $t$. The element $r_{t + 1}^{\left( 0 \right)}$  is the reward of the CS-DLMA node.  If the CS-DLMA node successfully transmitted a packet with length ${R_C}$ in time step $t$, then   $r_{t + 1}^{\left( 0 \right)} = {R_C} - H$; otherwise  $r_{t + 1}^{\left( 0 \right)} = 0$.  The element  $r_{t + 1}^{\left( i \right)}$,  $i = 1,2, \ldots ,L$,  is the reward of the node $i$  from other networks and $L$  is the total number of the nodes in other networks. If the node  $i$ has successfully transmitted a packet with length $R$  in time step $t$, then  $r_{t + 1}^{\left( i \right)} = R - H$; otherwise,  $r_{t + 1}^{\left( i \right)} = 0$.

\begin{figure*}[t]
	\normalsize
	\begin{IEEEeqnarray}{rCl}
		L\left( {\bm{\theta }} \right) = \frac{1}{{{N_E}  (L + 1)}}\sum\limits_{i = 0}^L {\sum\limits_{{e_\tau } \in E} {{{\left( {r_{\tau  + 1}^{\left( i \right)} + \gamma {Q^{\left( i \right)}}\left( {{s_{\tau  + 1}},{a_{\tau  + 1}};{{\bm{\theta }}^ - }} \right) - {Q^{\left( i \right)}}\left( {{s_\tau },{a_\tau };{\bm{\theta }}} \right)} \right)}^2}} }  \label{eqn5}
	\end{IEEEeqnarray}
	\vspace*{-0.15in}
\end{figure*}
\begin{figure*}[!t]
	\normalsize
	\begin{IEEEeqnarray}{rCl}
		{a_{\tau  + 1}} = \mathop {\arg \max }\limits_{a' \in \left\{ {0,1} \right\}} \sum\limits_{i = 0}^L {{f_\alpha }\left( {{Q^{\left( i \right)}}\left( {{s_{\tau  + 1}},a';{{\bm{\theta }}^ - }} \right)} \right)} \label{eqn6}
	\end{IEEEeqnarray}
	\vspace*{-0.15in}
\end{figure*}
\begin{figure*}[!t]
	\normalsize
	\begin{IEEEeqnarray}{rCl}
		L\left( {\bm{\theta }} \right) = \frac{1}{{{N_E}  (L + 1)}}\sum\limits_{i = 0}^L {\sum\limits_{{e_\tau } \in E} {{{\left( {\frac{{r_{\tau  + 1}^{\left( i \right)}}}{{d\left( {{a_\tau }} \right)}} \cdot \frac{{1 - {\gamma ^{d\left( {{a_\tau }} \right)}}}}{{1 - \gamma }} + {\gamma ^{d\left( {{a_\tau }} \right)}}{Q^{\left( i \right)}}\left( {{s_{\tau  + 1}},{a_{\tau  + 1}};{{\bm{\theta }}^ - }} \right) - {Q^{\left( i \right)}}\left( {{s_\tau },{a_\tau };{\bm{\theta }}} \right)} \right)}^2}} } \label{eqn7}
	\end{IEEEeqnarray}
	\vspace*{-0.15in}
\end{figure*}
\begin{figure*}[!t]
	\normalsize
	\begin{IEEEeqnarray}{rCl}
		{a_{t + 1}} = \left\{ {\begin{array}{*{20}{c}}
				{0,}&{{\rm{if }}\ {z_t} \ne IDLE,}\\
				{\mathop {\arg \max }\limits_{a' \in \left\{ {0,1, \ldots ,{R_{C\max }}} \right\}} \sum\limits_{i = 0}^L {{f_\alpha }\left( {{Q^{\left( i \right)}}\left( {{s_{t + 1}},a';{{\bm{\theta }}^ - }} \right)} \right),} }&{{\rm{if }}\ {z_t} = IDLE,{\rm{ with \ prob. }} \ 1 - \varepsilon ,}\\
				{{\rm{random \ choice \ in }}\left\{ {0,1, \ldots ,{R_{C\max }}} \right\},}&{{\rm{if }} \ {z_t} = IDLE,{\rm{ with \ prob. }} \ \varepsilon.}
		\end{array}} \right. \label{eqn8}
	\end{IEEEeqnarray}
	\hrulefill
\end{figure*}
\subsection{Non-Uniform Time-Step Multi-Dimensional DQN}
In our early work \cite{yu2019deep}, we put forth a multi-dimensional DQN framework for DLMA that deals with time steps of uniform duration. Specifically, in \cite{yu2019deep}, there was no carrier sensing  functionality for all involved MACs, and time-slotted systems with time slots of fixed duration were considered. The duration of a time slot in \cite{yu2019deep} corresponds to the duration of a packet transmission. By contrast, the current work extends the multi-dimensional DQN in \cite{yu2019deep} to scenarios in which the time slots for carrier sensing (i.e., minislots in this work) are of a smaller duration than the time slots for packet transmissions. For this extension, we need to modify the \textit{discounting mechanism} and \textit{action selection method} of conventional DQN. We lay out the principle for the modifications here.

In conventional DQN \cite{DQNpaper}, the outputs of the neural network are the approximated Q-values for different actions,  $\left\{ {Q\left( {s,a;{\bm{\theta }}} \right)|a \in \mathbb{A}} \right\}$, where the Q-value  $Q\left( {s,a;{\bm{\theta }}} \right)$ is the approximated cumulative discounted reward of a state-action pair  $\left( s,a\right) $. In the multi-dimensional DQN in \cite{yu2019deep}, the outputs of the neural network are a vector  $\left\{ {{Q^{\left( i \right)}}\left( {s,a;{\bm{\theta }}} \right)|a \in \left\{ {0,1} \right\},i \in \left\{ {0,1, \ldots ,L} \right\}} \right\}$, where ${Q^{\left( 0 \right)}}\left( {s,a;{\bm{\theta }}} \right)$ is the approximated cumulative discount reward of the DLMA node, ${Q^{\left( i \right)}}\left( {s,a;{\bm{\theta }}} \right)$,  $i = 1,2, \ldots ,L$, is the approximated cumulative discount reward of the node $i$  from other networks, and $a = 0$/$a = 1$ corresponds to ``NOT Transmit''/``Transmit'' (the number of actions in \cite{yu2019deep} is  two). Furthermore, the experience tuple ${e_t} = \left( {{s_t},{a_t},{r_{t + 1}},{s_{t + 1}}} \right)$ is augmented to  ${e_t} = \left( {{s_t},{a_t},{{\bm{r}}_{{\bm{t + 1}}}},{s_{t + 1}}} \right)$, wherein ${{\bm{r}}_{{\bm{t + 1}}}}$  is a vector consisting of the individual rewards of different nodes in the heterogeneous network, as opposed in the scalar reward ${r_{t + 1}}$ of a single entity in conventional DQN. With the above two modifications, in \cite{yu2019deep}, the loss function \eqref{eqn2} was rewritten as \eqref{eqn5}, and $a_{\tau + 1}$ in \eqref{eqn5} was chosen according to \eqref{eqn6}.



In this paper, for the study of CS-DLMA,  the time duration of each time step, i.e., the duration of each action  $a_t$, is non-uniform in that the packet length $R_C$  of the CS-DLMA node can be varying (recall that  ${a_t} \in \left\{ {0,1,2, \ldots ,{R_{C\max }}} \right\}$). The discounting mechanism in \eqref{eqn5} needs to be modified to take the non-uniform time-step into account. In particular, large time steps need to be discounted more than small time steps because the former extends further into the future.

To extend the uniform time-step multi-dimensional DQN in \cite{yu2019deep} to \textbf{non-uniform time-step multi-dimensional DQN}, the outputs of QNN are  modified to $\left\{ {{Q^{\left( i \right)}}\left( {s,a;{\bm{\theta }}} \right)|a \in \left\{ {0,1, \ldots ,{R_{C\max }}} \right\},i \in \left\{ {0,1, \ldots ,L} \right\}} \right\}$, i.e., the outputs of QNN are the approximated Q values for different actions and different nodes. In addition, we  let $d\left( {{a_t}} \right)$  denote the time duration of action  $a_t$ in terms of the number of minislots. Specifically,   $d\left( {{a_t}} \right) = 1$  if the CS-DLMA node performs carrier sensing over one minislot, and $d\left( {{a_t}} \right) = {R_C}$ if the CS-DSMA node transmits a packet of duration  ${R_C}$ minislots. We then augment the experience ${e_t} = \left( {{s_t},{a_t},{{\bm{r}}_{{\bm{t + 1}}}},{s_{t + 1}}} \right)$  to ${e_t} = \left( {{s_t},{a_t},d\left( {{a_t}} \right),{{\bm{r}}_{{\bm{t + 1}}}},{s_{t + 1}}} \right)$. Finally, the loss function \eqref{eqn5} can be modified  as \eqref{eqn7}.  Note that $ {a_{\tau  + 1}} $ in \eqref{eqn7} is the same as in \eqref{eqn6} except that the set of possible actions is $\left\{ {0,1, \ldots ,{R_{C\max }}} \right\}$ rather than $\left\{ {0,1} \right\}$.


We can write  $\frac{{r_{\tau  + 1}^{\left( i \right)}}}{{d\left( {{a_\tau }} \right)}} \cdot \frac{{1 - {\gamma ^{d\left( {{a_\tau }} \right)}}}}{{1 - \gamma }}$  in \eqref{eqn7} as  $\frac{{r_{\tau  + 1}^{\left( i \right)} \cdot (1 + \gamma  +  \cdots  + {\gamma ^{d\left( {{a_\tau }} \right) - 1}})}}{{d\left( {{a_\tau }} \right)}}$, corresponding to amortizing the reward $r_{\tau  + 1}^{\left( i \right)}$  in a non-uniform time-step over   minislots by minislot discounting. Now, the training of non-uniform time-step multi-dimensional DQN can be done by minimizing the loss function \eqref{eqn7} using Stochastic Gradient Descent \cite{lecun2015deep}. 

For action selection in CS-DLMA, we put forth a \textbf{carrier-sense $\varepsilon$-greedy} algorithm. Suppose that at the beginning of time step  $t+1$, the state of the CS-DLMA node is ${s_{t + 1}}$  and the CS-DLMA node needs to select an action  ${a_{t + 1}}$. The carrier-sense  $\varepsilon$-greedy algorithm that decides ${a_{t + 1}}$  is given by \eqref{eqn8}.


\begin{figure}[!t]
	\centering
	\includegraphics[scale=0.65]{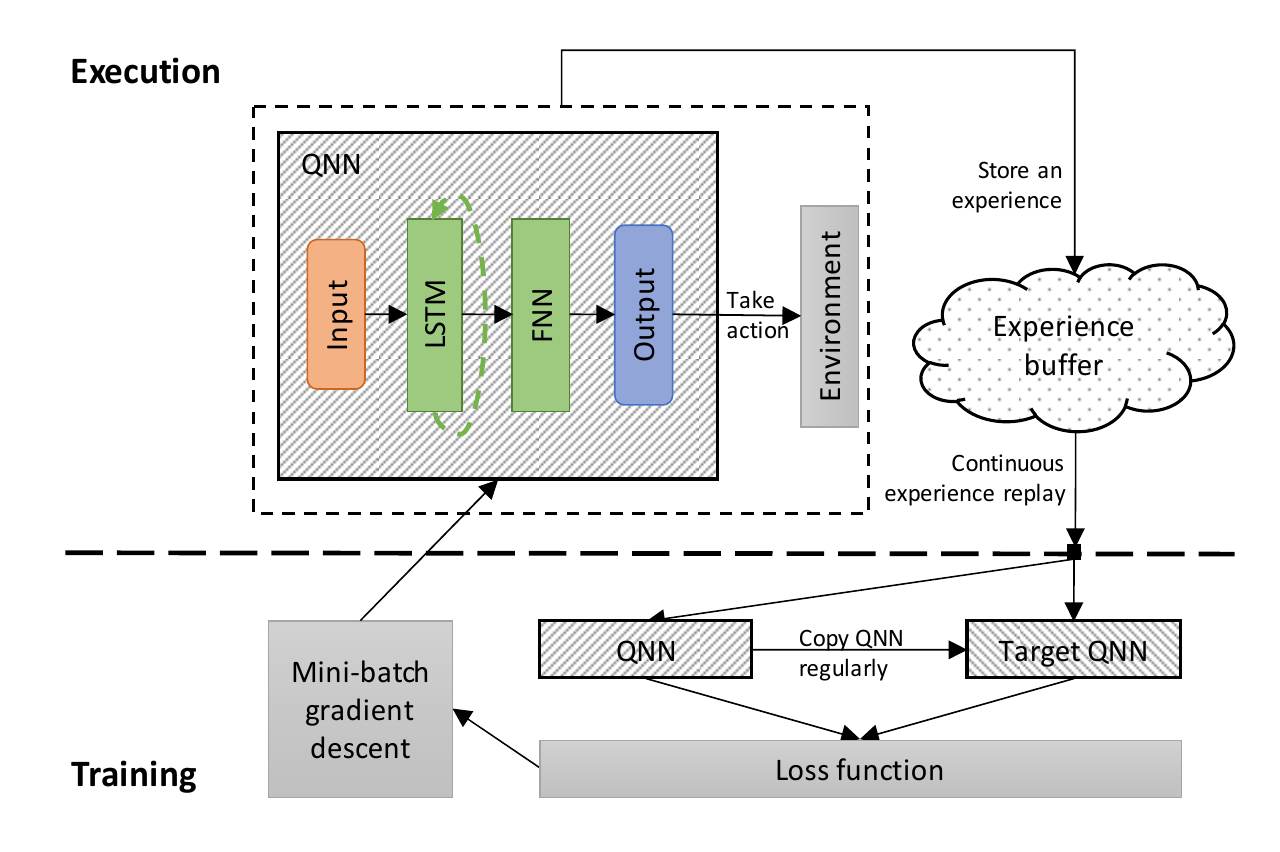}
	\caption[]{DQN architecture for realizing  CS-DLMA.}
	\label{fig:architecture}
\end{figure} 

We now explain \eqref{eqn8} line by line. The first line in \eqref{eqn8} is a result of the carrier sensing mechanism. A node operating carrier sensing needs to sense the network to be idle before it can transmit. The sensing operation will take a certain amount of time. For our system, we assume one minislot is used for sensing (since even if sensing can be completed in less than one minislot, the node will still need to wait for the next minislot to begin transmission, if the wireless channel is sensed to be idle). The first line in \eqref{eqn8} is attributed to the non-zero amount of time (i.e., one minislot in our case) needed for carrier sensing. If the channel is not idle in time step  $t$, then the CS-DLMA node cannot transmit in time step   $t+1$. In time step  $t$, the channel could be non-idle either because the CS-DLMA node was transmitting or another node is transmitting and the CS-DLMA node sensed the medium to be busy, i.e.  ${z_t} =$ \textit{BUSY}, \textit{SUCCESSFUL} or \textit{COLLIDED}.  

The second line and third line in \eqref{eqn8} describes action selection in time step  $t+1$ if the CS-DLMA node did not transmit at time step   $t$, and it sensed the medium to be idle at time step  $t$ (i.e., other nodes did not transmit either). In this case, the CS-DLMA node can decide to transmit or not to transmit in time step  $t+1$. If it decides to transmit, the CS-DLMA node also needs to decide the packet length of the transmission. With an    $\varepsilon$-greedy algorithm, with probability   $\varepsilon$ the choice is made uniform randomly, as in the third line of \eqref{eqn8}. The second line of \eqref{eqn8} is a departure from the conventional  $\varepsilon$-greedy DQN algorithm. In conventional DQN, with probability $1 - \varepsilon$  the action that yields the maximum Q value $\left\{ {Q\left( {s,a;{\bm{\theta }}} \right)|a \in A} \right\}$ is selected \cite{DQNpaper}. To capture the essence of  $\alpha$-fairness objective, our multi-dimensional DQN selects the action that maximizes an   $\alpha$-fairness nonlinear combination of different Q values.

\subsection{CS-DLMA Implementation}\label{sec:implementation}
Fig. \ref{fig:architecture} shows the overall DQN architecture that realizes CS-DLMA.\footnote{The simulation codes of CS-DLMA are partly open-sourced: https://github.com/YidingYu/CS-DLMA.} We now describe three key components in the architecture: 1) neural network, 2) experience buffer, and 3) continuous experience replay.
\subsubsection{Neural Network} The neural network, i.e., QNN, used in non-uniform multi-dimensional DQN is a recurrent neural network (RNN). The RNN consists of an input layer, two hidden layers, and an output layer. The input to the RNN is the current state. The two hidden layers consist of a long-short-term-memory (LSTM) \cite{hochreiter1997long} layer and a feedforward layer. The outputs are the approximated Q values for different actions and different nodes given the input state. 
\begin{figure}[!t]
	\centering
	\subfigure[{\scriptsize FNN}]{
		\label{fig:FNN}
		\begin{minipage}[t]{0.1\textwidth}
			\centering
			\includegraphics[scale=0.7]{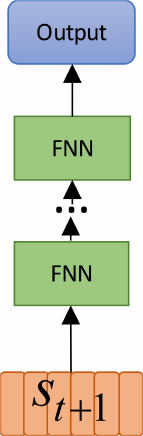}
			\vspace{-0.05in}
	\end{minipage}}
	\hspace{-1.8cm}
	\subfigure[{\scriptsize RNN}]{
		\label{fig:RNN}
		\begin{minipage}[t]{0.45\textwidth}
			\centering
			\includegraphics[scale=0.7]{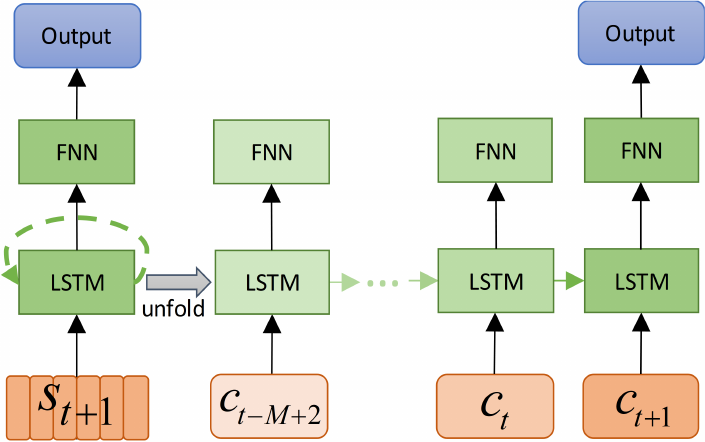}
	\end{minipage}}
	\hspace{-2cm}
	\caption{{\scriptsize FNN-based QNN versus RNN-based QNN.} }
	\label{fig:FNN_RNN}
\end{figure}

Instead of RNN, a feedforward neural network (FNN) could also be used, wherein the hidden layers are all pure feedforward layers. Fig. \ref{fig:FNN_RNN} shows the difference between the FNN-based QNN and the  RNN-based QNN in processing ${s_{t + 1}} = \left[ {{c_{t - M + 2}}, \cdots ,{c_t},{c_{t + 1}}} \right]$  received from the input layer at time step  $t+1$. After receiving  ${s_{t + 1}}$, FNN processes it directly; by contrast, after receiving  ${s_{t + 1}}$, RNN processes the elements,  ${c_{t - M + 2}}, \cdots ,{c_t},{c_{t + 1}}$, in   ${s_{t + 1}}$  sequentially, keeping an internal state as it injects the elements one by one into the input in a sequential manner. In this way, the causal relationship between the elements in  ${s_{t + 1}}$  (e.g.,  $c_t$  precedes  ${c_{t + 1}}$) is explicitly embedded in the way RNN processes the input \cite{hochreiter1997long}. On the other hand, the causal relationship between elements in  ${s_{t + 1}}$ is not explicitly given to FNN. FNN will need to learn this relationship if it manages to learn at all.

\subsubsection{Experience Buffer} For implementation, it is inefficient to store experiences in the form of ${e_t} = \left( {{s_t},{a_t},d\left( {{a_t}} \right),{{\bm{r}}_{{\bm{t + 1}}}},{s_{t + 1}}} \right)$  since two consecutive experiences have many  common elements. For example,  ${s_{t + 1}}$ in ${e_t}$ is only a time-shifted version of ${s_t}$   in ${e_t}$   with the headend discarded and a new tailend appended. It is superfluous to store the overlapped elements for both experiences. A more efficient implementation is to store the abbreviated experience  $\left( {{c_t},{a_t},d\left( {{a_t}} \right),{{\bm{r}}_{{\bm{t + 1}}}},{c_{t + 1}}} \right)$. The complete experience  ${e_t}$   can be obtained from consecutive abbreviated experiences by means of \textit{continuous experience replay}.

\subsubsection{Continuous Experience Replay} 
\begin{figure}[!t]
	\centering
	\includegraphics[scale=1]{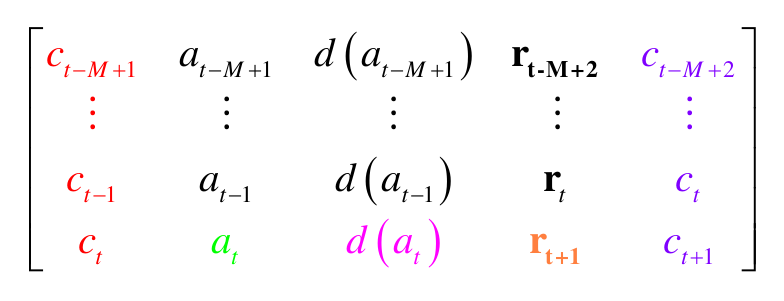}
	\caption[]{A sample in continuous experience replay.}
	\label{fig:sample}
\end{figure}

In conventional experience replay \cite{DQNpaper}, random experiences are sampled from the experience buffer to compute the loss function, with each sample being an experience  ${e_t} = \left( {{s_t},{a_t},d\left( {{a_t}} \right),{{\bm{r}}_{{\bm{t + 1}}}},{s_{t + 1}}} \right)$. After downsizing the experience to  $\left( {{c_t},{a_t},d\left( {{a_t}} \right),{{\bm{r}}_{{\bm{t + 1}}}},{c_{t + 1}}} \right)$, we will sample continuous experiences instead to extract the information necessary for computing the loss function \eqref{eqn7}. As illustrated in Fig. \ref{fig:sample}, each sample contains  $M$ continuous experiences, and we can extract  ${s_t} = \left[ {{c_{t - M + 1}}, \cdots ,{c_{t - 1}},{c_t}} \right]$,  ${a_t}$,  $d\left( {{a_t}} \right)$,  ${{\bm{r}}_{{\bm{t + 1}}}}$, ${s_{t + 1}} = \left[ {{c_{t - M + 2}}, \cdots ,{c_t},{c_{t + 1}}} \right]$ from this sample.

\section{Performance Evaluation}

This section evaluates the performance of CS-DLMA.  After introducing the simulation setup, we first investigate the coexistence of CS-DLMA with TDMA and ALOHA, two MAC protocols without carrier sensing. Following that, we investigate the coexistence of CS-DLMA with WiFi, a MAC protocol with carrier sensing. For concreteness,  this paper focuses on saturated networks, i.e., all the nodes in the networks always have packets to transmit. In addition, since we have no control of TDMA, ALOHA and WiFi, we assume the packet lengths of these nodes are fixed in our evaluation.

\subsection{Simulation Setup}\label{sec:set_up}
\subsubsection{Hyperparameters} We adopt the RNN architecture in CS-DLMA unless stated otherwise (we will show our motivation to use RNN by comparing the performance of RNN with FNN in Appendix \ref{sec:appendix}).  As shown in Fig. \ref{fig:architecture}, the RNN architecture has two hidden layers: one LSTM layer followed by one feedforward layer. The number of neurons for each layer is 64 and the activation functions are ReLU \cite{lecun2015deep}. Since we assume CS-DLMA does not know the mechanisms of the coexisting MACs, we use a relatively large state history length  $M$  to cover a longer history so as to learn the behavior of potentially complex MACs. Specifically, for our simulations, we set  $M = 20$, i.e., the state of CS-DLMA covers the action-observation pairs in the past 20 time steps.  The value of  $\varepsilon$ in the carrier-sense  $\varepsilon$-greedy algorithm is initially set to 1 and decreases at a rate of 0.995 every time step until its value reaches 0.005, i.e.,  $\varepsilon$ is updated by $\varepsilon  \leftarrow \max \left\{ {0.995*\varepsilon ,0.05} \right\}$  in each time step. The discount factor  $\gamma$  in \eqref{eqn7} is set to 0.999. The size of the experience buffer is 1000 and the experience buffer is updated in a FIFO manner \cite{yu2019deep}. The RMSProp algorithm \cite{tieleman2012lecture} is used to conduct minibatch gradient descent over the loss function \eqref{eqn7}.  The minibatch size  ${N_E}$ is set to 32.  The target network is updated every 20 time steps. Table \ref{table:hyperparameters} summarizes the values of the hyperparameters. 

\begin{table}[htbp]
	\centering
	\renewcommand\arraystretch{1.2}
	\caption{\textcolor{black}{CS-DLMA Hyperparameters}}
	\label{table:hyperparameters}
	\begin{tabular}{|l|l|}
		\hline
		\textbf{Hyperparameter} & \textbf{Value} \\ \hline
		State history length  $ M $& 20\\ \hline
		$ \varepsilon $ in carrier-sense $ \varepsilon $-greedy algorithm & 1 to 0.005           \\ \hline
		Discount factor $ \gamma $ & 0.999\\ \hline
		Experience buffer size & 1000\\ \hline
		Experience-replay minibatch size $ N_E $ & 32\\ \hline
		Target network update frequency  & 20\\  \hline
	\end{tabular}
\end{table}

\begin{figure*}[t]
	\centering
	\includegraphics[scale=0.75]{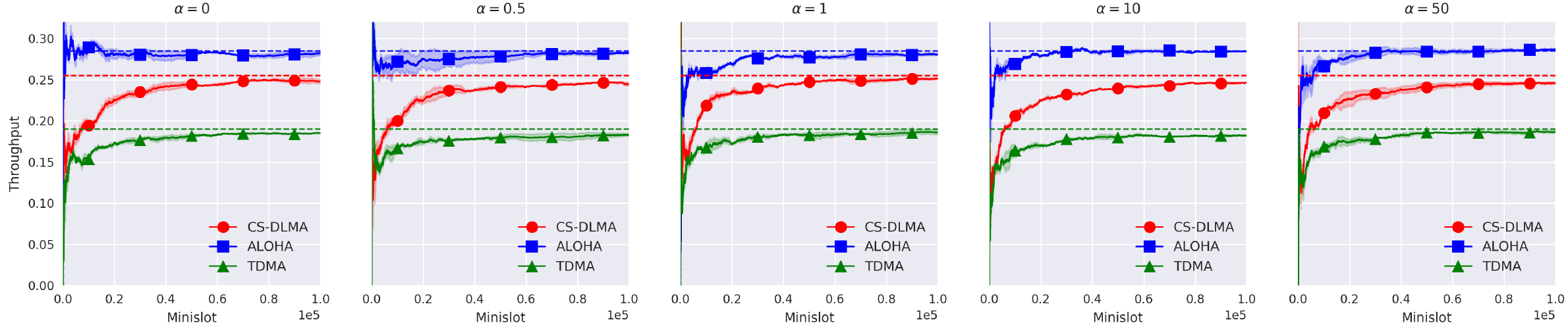}
	\caption[]{Individual throughputs of CS-DLMA, TDMA, and ALOHA for different  $\alpha$ values. The solid lines are the throughputs of individual nodes. Each solid line is averaged over 10 different runs, with the shaded areas being areas within the standard deviation. The dashed lines are for benchmarking purposes. They are the node throughputs when the CS-DLMA node is replaced by a model-aware node operating a model-aware optimal strategy. }
	\label{fig:eva_RNN}
\end{figure*}
\subsubsection{Performance Metric} We evaluate the performance of CS-DLMA by examining whether the objective in \eqref{eqn4} can be achieved. In particular, we define the ``\textbf{throughput}'' of node  $i$ at time step  $t$ by 
\begin{equation}
\sum\nolimits_{t' = 0}^t {r_{t' + 1}^{\left( i \right)}} /\sum\nolimits_{t' = 0}^t {d\left( {{a_{t'}}} \right)}
\end{equation}
where  $r_{t' + 1}^{\left( i \right)}$ is the reward of node  $i$   at the end of time step  $t'$ and $d\left( {{a_{t'}}} \right)$  is the time duration of action  ${a_{t'}}$ in terms of number of minislots.  The throughput here is the average reward and reflects the performance of each node in the long run.
\subsection{CS-DLMA coexists with TDMA and ALOHA}\label{sec:coexist_TDMA_ALOAH}
This subsection investigates the coexistence of one CS-DLMA node with one TDMA node and one ALOHA node. We first introduce the settings of each node. We then examine if CS-DLMA can achieve a general  $\alpha$-fairness objective when coexisting with TDMA and ALOHA. 
\begin{figure*}[!t]
	\centering
	\includegraphics[scale=0.75]{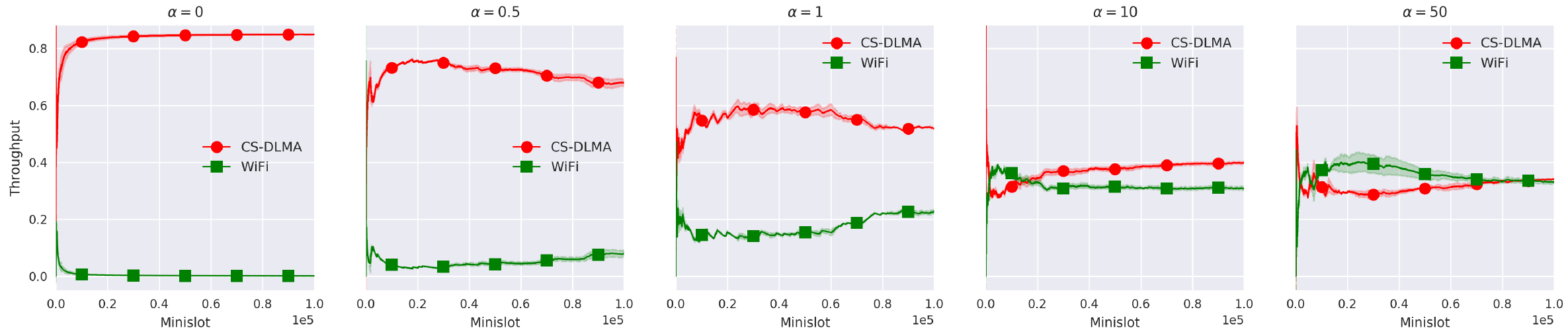}
	\caption[]{Individual throughputs of CS-DLMA and WiFi for different $\alpha$ values. Each line is averaged over 10 different runs, with the shaded areas being areas within the standard deviation.}
	\label{fig:eva_wifi1}
\end{figure*}

In our experimental setup, the TDMA node occupies the second and the fifth TDMA slots within a TDMA frame of five TDMA slots;  the ALOHA node transmits with a fixed probability of $q = 0.5$  in each ALOHA slot. The packet lengths of TDMA and ALOHA are both fixed at 10 minislots. The CS-DLMA node practices our CS-DLMA protocol and can transmit packets of variable length, with a maximum length of 10 minislots. For benchmarking, in place of the model-free CS-DLMA node, we imagine a model-aware node that is aware of the packet length as well as the MAC mechanisms of TDMA and ALOHA. As for CS-DLMA, we also assume that the packet length of the model-aware node can vary from 1 to 10 minislots. The optimal strategy of the model-aware node summarized below can achieve the general  $\alpha$-fairness objective:

\textit{At the beginning of each TDMA/ALOHA slot, the model-aware node performs carrier sensing. If the channel is idle, the model-aware node transmits in the next 9 minislots; if the channel is busy (either TDMA or ALOHA transmits), the model-aware node keeps silent in the next 9 minislots. }

A point to note here is that the optimal strategies of the model-aware node are the same for different  values. The detailed analyses are provided in Appendix \ref{sec:appendix-B}.

We now examine if CS-DLMA can manage to find the optimal strategies for different   $\alpha$ values without being aware of the MACs of TDMA and ALOHA.  Fig. \ref{fig:eva_RNN} plots the individual throughputs of CS-DLMA, TMDA and ALOHA achieved by CS-DLMA as well as the corresponding optimal individual throughputs achieved by the model-aware node. As can be seen from Fig. \ref{fig:eva_RNN}, for different  $\alpha$  values, the individual throughputs of each node all approximate their corresponding optimal results, indicating that CS-DLMA indeed can find a strategy that achieves   $\alpha$-fairness objectives of different $\alpha$ values. 

\subsection{CS-DLMA coexists with WiFi}\label{sec:coexist_wifi}
We next investigate the coexistence of one CS-DLMA node with one WiFi node. The CS-DLMA node is the same as in Section \ref{sec:coexist_TDMA_ALOAH}. The WiFi node uses the following settings: the packet length is fixed at 10 minislots; the initial window size is 2; the maximum backoff stage is 6.

We first present the individual throughputs of CS-DLMA and WiFi for different $\alpha$ values. As can be seen from Fig. \ref{fig:eva_wifi1}, when the value of $\alpha$  increases from 0 to 50, the throughputs of CS-DLMA and WiFi get closer. In particular, when $\alpha = 0$, CS-DLMA aims to maximize the sum throughput and the strategy found by CS-DLMA is a greedy strategy, i.e., CS-DLMA always transmits if the channel is sensed idle; when $\alpha$  increases, CS-DLMA becomes less aggressive and leaves more opportunities for WiFi until the throughput of CS-DLMA and WiFi are almost equal. This demonstrates that CS-DLMA indeed can adjust its strategy according to the value of  $\alpha$.  

For comparison purposes, we replace CS-DLMA with p-persistent CSMA (p-CSMA) \cite{Tanenbaum:2010:CN:1942194} in the above experiment, i.e., we consider the coexistence of p-CSMA with WiFi. If the channel is sensed idle, p-CSMA transmits a packet with a probability of  $p$ ($p \in \left[ {0,1} \right]$). The value of  $p$ can be adjusted to achieve different throughput allocations between p-CSMA and WiFi when they coexist. 
\begin{figure}[!t]
	\centering
	\includegraphics[scale=0.32]{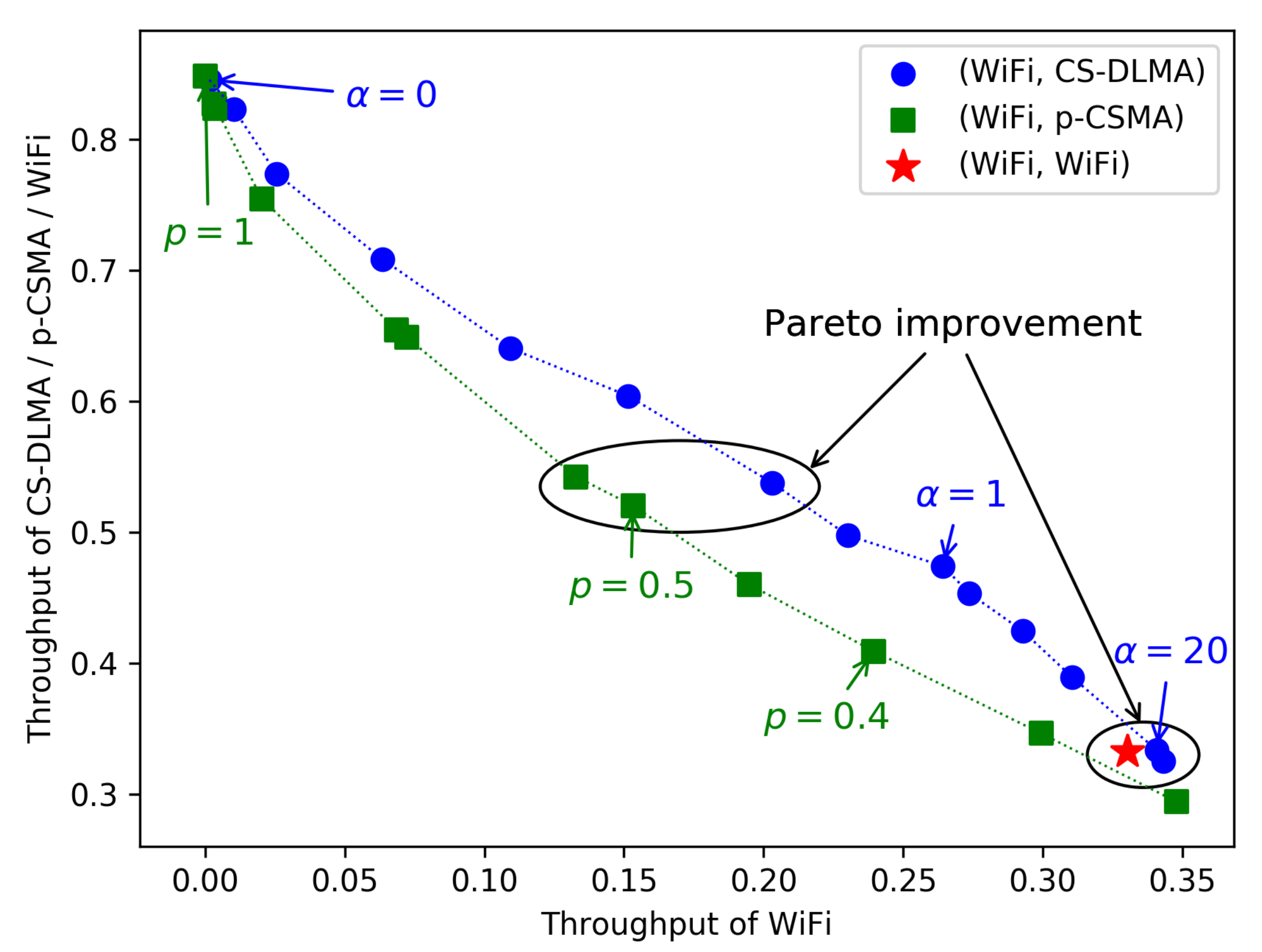}
	\caption[]{Individual throughputs of CS-DLMA/p-CSMA and WiFi at different  $\alpha$/$p$  values. The red star corresponds to the individual throughputs of two homogeneous WiFi nodes.}
	\label{fig:eva_wifi2}
\end{figure}

Fig. \ref{fig:eva_wifi2} plots the throughputs of CS-DLMA/p-CSMA versus WiFi. Specifically, in Fig. \ref{fig:eva_wifi2}, the x-axis is the throughput of WiFi and the y-axis is the throughput of CS-DLMA/p-CSMA.  Each circle corresponds to the throughputs allocation achieved by CS-DLMA with a particular  $\alpha$; each square corresponds to the throughputs allocation achieved by p-CSMA with a particular  $p$. As can be seen from Fig. \ref{fig:eva_wifi2}, CS-DLMA can achieve Pareto improvement \cite{myerson2013game} over p-CSMA when coexisting with WiFi. Interestingly, if we also plot the individual throughputs of two homogeneous WiFi nodes---denoted by the red star in Fig. \ref{fig:eva_wifi2}---we find that CS-DLMA can also achieve Pareto improvement over WiFi when coexisting with WiFi.  

An intuitive reason why our CS-DLMA manages to obtain performance more Pareto efficiently than p-CSMA is that the CS-DLMA node looks at a longer state history $M$  before making a decision on the action to follow while the p-CSMA node does not (its effective  $M$ is 1). Since the behavior of the WiFi node is not Markovian in that its behavior does not depend just on whether it is currently transmitting or not, but also on its experiences stretching further to the past, having access to a longer state history will help.

\section{Multi-Node CS-DLMA Framework}\label{sec:multi-node}
Section \ref{sec:one-node} introduced the CS-DLMA framework with only one CS-DLMA node. This section generalizes the one-node CS-DLMA framework to the multi-node CS-DLMA framework. With this framework, we will investigate the coexistence of multiple CS-DLMA nodes with multiple other nodes. 

Revisiting the system model introduced in Section \ref{sec:system_model}, we know that although the CS-DLMA network has no control of other networks, CS-DLMA nodes running the same protocol can be coordinated.  For the multi-node CS-DLMA framework studied here, we put forth a CS-DLMA protocol to enable CS-DLMA network to achieve the  $\alpha$-fairness objective. In particular, we assume there is a CS-DLMA gateway associated with the CS-DLMA nodes in the CS-DLMA network. The gateway is responsible for coordinating the operations of the CS-DLMA nodes so that they coexist among themselves and coexist with nodes running other protocols to meet the  $\alpha$-fairness objective.
\begin{figure*}[!t]
	\normalsize
	\begin{IEEEeqnarray}{rCl}
		L\left( {\bm{\theta }} \right) = \frac{1}{{{N_E} \left( {L + 1} \right)}}\sum\limits_{i = 0}^L {\sum\limits_{{e_\tau } \in E} {{{\left( {\frac{{r_{\tau  + 1}^{\left( i \right)}}}{{d\left( {{a_\tau }} \right)}} \cdot \frac{{1 - {\gamma ^{d\left( {{a_\tau }} \right)}}}}{{1 - \gamma }} + {\gamma ^{d\left( {{a_\tau }} \right)}}{Q^{\left( i \right)}}\left( {{s_{\tau  + 1}},{a_{\tau  + 1}};{{\bm{\theta }}^ - }} \right) - {Q^{\left( i \right)}}\left( {{s_\tau },{a_\tau };{\bm{\theta }}} \right)} \right)}^2}} } \label{eqn10}
	\end{IEEEeqnarray}
	\vspace*{-0.15in}
\end{figure*}
\begin{figure*}[!t]
	\normalsize
	\begin{IEEEeqnarray}{rCl}
		{a_{\tau  + 1}} = \mathop {\arg \max }\limits_{a' \in \left\{ {0,1, \ldots ,{R_{C\max }}} \right\}} \left\{ {\left( {N - L} \right) {f_\alpha }\left( {\frac{{{Q^{\left( i \right)}}\left( {{s_{\tau  + 1}},a';{{\bm{\theta }}^ - }} \right)}}{{N - L}}} \right) + \sum\limits_{i = 1}^L {{f_\alpha }\left( {{Q^{\left( i \right)}}\left( {{s_{\tau  + 1}},a';{{\bm{\theta }}^ - }} \right)} \right)} } \right\}\label{eqn11}
	\end{IEEEeqnarray}
	\vspace*{-0.15in}
\end{figure*}

\begin{figure*}[!t]
	\normalsize
	\begin{IEEEeqnarray}{rCl}
		{\small 		{a_{t + 1}} = \left\{ {\begin{array}{*{20}{c}}
					{0,}&{{\rm{if }}\ {z_t} \ne IDLE,}\\
					{\mathop {\arg \max }\limits_{a' \in \left\{ {0,1, \ldots ,{R_{C\max }}} \right\}} \left( {N - L} \right)  {f_\alpha }\left( {\frac{{{Q^{\left( i \right)}}\left( {{s_{t + 1}},a';{{\bm{\theta }}^ - }} \right)}}{{N - L}}} \right) + \sum\limits_{i = 1}^L {{f_\alpha }\left( {{Q^{\left( i \right)}}\left( {{s_{t + 1}},a';{{\bm{\theta }}^ - }} \right)} \right),} }&{{\rm{if }}\ {z_t} = IDLE,{\rm{ with \ prob. }}\ 1 - \varepsilon ,}\\
					{{\rm{random \ choice \ in }}\left\{ {0,1, \ldots ,{R_{C\max }}} \right\},}&{{\rm{if }}\ {z_t}= IDLE,{\rm{ with \ prob. }}\ \varepsilon .}\label{eqn12}
			\end{array}} \right.}
	\end{IEEEeqnarray}
	\hrulefill
\end{figure*}

If the CS-DLMA gateway decides to perform carrier sensing, it will listen to the channel and check whether the channel is occupied by the nodes from other networks; if the CS-DLMA gateway decides to transmit a packet, it will select one of the CS-DLMA nodes in a round-robin manner to transmit (the CS-DLMA gateway itself is also a CS-DLMA node). The instruction from the CS-DLMA gateway to the other CS-DLMA nodes can be sent through a control channel within the CS-DLMA network. For example, the control channel can be implemented as a ``short time slot'' before each packet transmission. The time duration of the ``short time slot'' can be even smaller than a minislot and can be neglected in the performance evaluation.\footnote{For concreteness and for simplicity, we focus on a design with centralized coordination of all DL-CSMA nodes by a gateway here. Decentralized coordination is also possible. For example, if all the CS-DLMA nodes run the same algorithm as the gateway algorithm described in this paper, and all CS-DLMA nodes have the same observations,  then  the CS-DLMA nodes will be in consensus as to the action to be taken by the CS-DLMA network next (i.e., whether a CS-DLMA node should transmit and if so, which CS-DLMA node should transmit).  For the decentralized implementation, there will be no need for a control channel for a central controller (gateway) to send instructions to the CD-DLMA nodes. However, how to ensure consensus among the CS-DLMA nodes, taking into consideration the possibility of discrepancies in their observations, will be a key issue.}

\begin{figure*}[!t]
	\centering
	\includegraphics[scale=0.75]{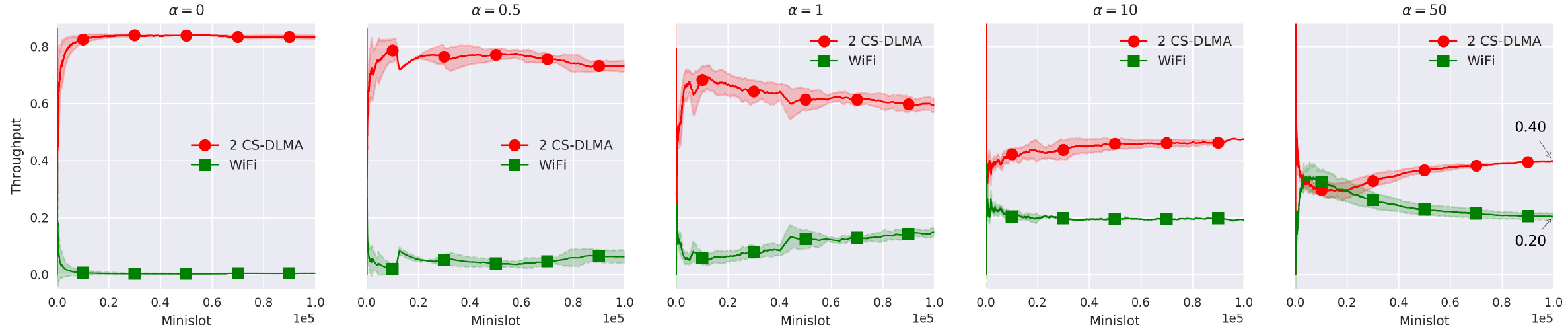}
	\caption[]{Sum throughputs of two CS-DLMA nodes, and the throughput of one WiFi node for different $\alpha$  values. Each line is averaged over 10 different runs, with the shaded areas being areas within the standard deviation.}
	\label{fig:eva_wifi3}
\end{figure*}

We now transform the multiple access problem faced by the CS-DLMA network to a reinforcement learning problem. In particular, our multi-node CS-DLMA framework is the same as the one-node CS-DLMA framework except that the following modifications are made:
\subsubsection{Action} At the beginning of each time step  $t$, the CS-DLMA gateway decides an action  ${a_t} \in \left\{ {0,1,2, \ldots ,{R_{C\max }}} \right\}$. If  ${a_t} = 0$, the CS-DLMA gateway will perform carrier sensing in the next minislot. After that, it will get an observation    ${z_t} = $ \textit{BUSY} or \textit{IDLE}, indicating whether the channel is being occupied or not occupied by other nodes. If  ${a_t} = {R_C} \in \left\{ {1,2, \ldots ,{R_{C\max }}} \right\}$, the CS-DLMA node will select one CS-DLMA node in a round-robin manner to transmit a packet with a length of  ${R_C}$ in the next  ${R_C}$ minislots. After that, it will get an observation ${z_t} = $ \textit{SUCCESSFUL} or \textit{COLLIDED}, indicating whether the packet is successfully received or not.
\subsubsection{Reward} After taking action  ${a_t}$, the CS-DLMA gateway obtains a reward vector ${{\bm{r}}_{{\bm{t + 1}}}} = \left[ {r_{t + 1}^{\left( 0 \right)},r_{t + 1}^{\left( 1 \right)},r_{t + 1}^{\left( 2 \right)}, \ldots ,r_{t + 1}^{\left( L \right)}} \right]$  from the environment at the end of time step  $t$. The element $r_{t + 1}^{\left( 0 \right)}$ is the reward of the CS-DLMA network.  If any CS-DLMA node successfully transmitted a packet with length ${R_C}$  in time step $t$, then   $r_{t + 1}^{\left( 0 \right)} = {R_C} - H$; otherwise  $r_{t + 1}^{\left( 0 \right)} = 0$.  The reward of the node $i$  from other networks,   $r_{t + 1}^{\left( i \right)}$, $i = 1,2, \ldots ,L$,  has the same definition as in Section \ref{sec:action_state_reward}. 
\subsubsection{Non-Uniform Time-Step Multi-Dimensional DQN} The outputs of the neural network in non-uniform multi-dimensional DQN are still denoted by  $\left\{ {{Q^{\left( i \right)}}\left( {s,a;{\bm{\theta }}} \right)|a \in \left\{ {0,1, \ldots ,{R_{\max }}} \right\},i \in \left\{ {0,1, \ldots ,L} \right\}} \right\}$, but ${Q^{\left( 0 \right)}}\left( {s,a;{\bm{\theta }}} \right)$ here is the approximated cumulative discounted reward of the CS-DLMA network, rather than the approximated cumulative discount reward of one particular CS-DLMA node (this modification is consistent with the definition of reward for multi-node CS-DLMA). The loss function is now given by \eqref{eqn10}.

Note that \eqref{eqn10} has the same form as \eqref{eqn7}, but the index $i=0$  refers to the CS-DLMA network rather than a particular CS-DLMA node. In addition, in \eqref{eqn10}, ${a_{\tau  + 1}}$  is different from \eqref{eqn6}, but is given by \eqref{eqn11}.

The first term $\left( {N - L} \right) \cdot {f_\alpha }\left( {{Q^{\left( i \right)}}\left( {{s_{\tau  + 1}},a';{{\bm{\theta }}^ - }} \right)/\left( {N - L} \right)} \right)$  in \eqref{eqn11}  is the utility function of the CS-DLMA network, where  $N - L$ is the number of all the CS-DLMA nodes and ${Q^{\left( i \right)}}\left( {{s_{\tau  + 1}},a';{{\bm{\theta }}^ - }} \right)/\left( {N - L} \right)$  can be regarded as the approximated cumulative discounted reward of each CS-DLMA node. The second term $\sum\nolimits_{i = 1}^L {{f_\alpha }\left( {{Q^{\left( i \right)}}\left( {{s_{\tau  + 1}},a';{{\bm{\theta }}^ - }} \right)} \right)}$ is the sum of utility functions of all the nodes from other networks. 
\subsubsection{Carrier-Sense  $\varepsilon$-greedy Algorithm} The action selection method in \eqref{eqn8} should also be modified for the multi-node CS-DLMA framework. In particular, in time step  $t+1$, the action ${a_{t + 1}}$  is given by  \eqref{eqn12}.

\section{Multi-Node CS-DLMA Performance Evaluation}
This section evaluates the performance of the multi-node CS-DLMA framework. We first consider the coexistence of two CS-DLMA nodes with one WiFi node to examine if our multi-node CS-DLMA framework can adjust its transmission strategy according to both the value of $\alpha$  and the number of CS-DLMA nodes. One of the two CS-DLMA nodes is designated as the gateway.  As in Sections \ref{sec:coexist_TDMA_ALOAH} and \ref{sec:coexist_wifi}, we also assume CS-DLMA nodes can transmit packets of variable length, with a maximum length of 10 minislots. The settings of the WiFi node is the same as in Section \ref{sec:coexist_wifi}. 

Fig. \ref{fig:eva_wifi3} plots the sum throughput of the two CS-DLMA nodes and the throughput of the WiFi node.  As can be seen from Fig. \ref{fig:eva_wifi3}, when $\alpha$  increases, the sum throughput of CS-DLMA and the throughput of WiFi get closer.  Specifically, when  $\alpha  = 50$, the sum throughput of CS-DLMA is twice the throughput of WiFi ($0.40/0.20 = 2$), which means the throughput of each CS-DLMA node is equal to the throughput of WiFi. This is consistent with our observation in Fig. \ref{fig:eva_wifi1} that when  $\alpha  = 50$, the throughput of one CS-DLMA node is equal to the throughput of one WiFi node. This demonstrates that our formulation of multi-node CS-DLMA can adjust the weight of CS-DLMA according to the number of CS-DLMA nodes. 

To further demonstrate the performance of the multi-node CS-DLMA framework, we now consider three coexistence scenarios:
\begin{enumerate}
	\item four CS-DLMA nodes with four WiFi nodes;
	\item four p-CSMA nodes with four WiFi nodes;
	\item eight WiFi nodes. 
\end{enumerate}

In scenario 1), the value $\alpha$  is set to 50, i.e., we want to achieve equal throughputs among four CS-DLMA nodes and four WiFi nodes; in scenario  2), each p-CSMA node adopts the same value of  $p$, and we adjust the value of  $p$  to let the throughput of each p-CSMA node be equal to the throughput of each WiFi node; in scenario 3), eight WiFi nodes are homogeneous. In addition, CS-DLMA, p-CSMA, and WiFi all adopt the same settings as in Section \ref{sec:coexist_wifi}. 

\begin{figure}[!t]
	\centering
	\includegraphics[scale=0.32]{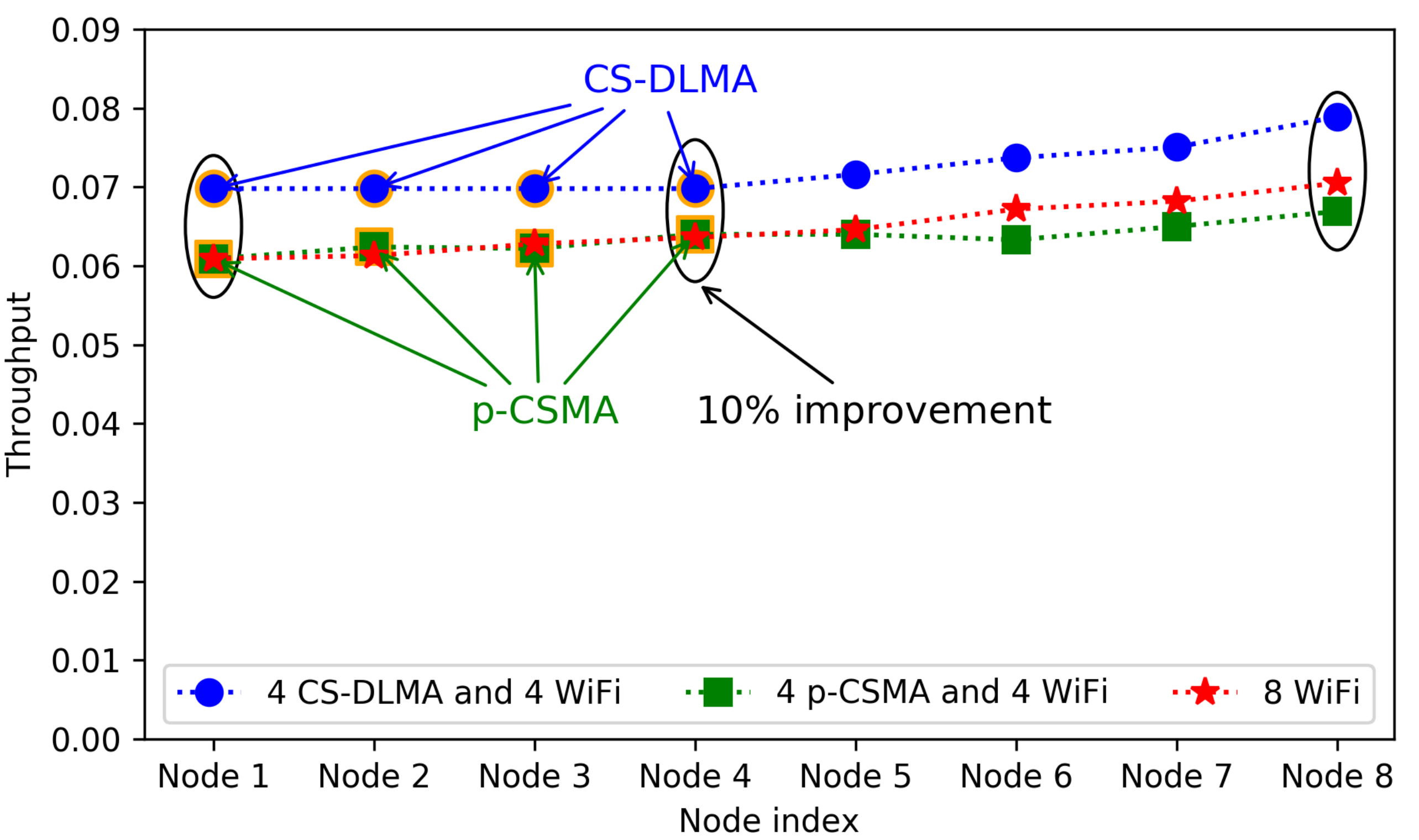}
	\caption[]{Individual throughput of each node in three coexistence scenarios: 1) four CS-DLMA nodes ($\alpha=50$) with four WiFi nodes, 2) four p-CSMA nodes ($p=0.124$) with four WiFi nodes and 3) eight WiFi nodes.}
	\label{fig:eva_wifi4}
\end{figure} 
Fig. \ref{fig:eva_wifi4} presents the individual throughputs of each node in the above three scenarios. Overall, roughly equal throughputs among all nodes can be achieved in all scenarios. However, the throughput in scenario 1) is about $10\%$ higher than those of scenarios 2) and 3). 

\section{Conclusion}
In this paper, we developed a deep reinforcement learning multiple access protocol with carrier sensing capability, referred to as CS-DLMA. The goal of CS-DLMA is to enable efficient and equitable spectrum sharing among a group of co-located heterogeneous wireless networks. A salient feature of CS-DLMA is that it can coexist harmoniously with other MAC protocols in the heterogeneous environment without knowing the MAC details of other networks. In particular, we demonstrated that CS-DLMA can achieve a general  $\alpha$-fairness objective \cite{mo2000fair} when coexisting with TDMA, ALOHA, and WiFi protocols by adjusting its own transmission strategies. Interestingly, we also found that CS-DLMA is more Pareto efficient than other CSMA protocols, e.g., p-persistent CSMA, when coexisting with WiFi.

The underpinning DRL technique in CS-DLMA is deep Q-network (DQN). However, the original DQN and its extension multi-dimensional DQN \cite{yu2019deep} are not applicable for CSMA protocols design due to the underlying uniform time-step assumption in the DQN framework---for CSMA protocols,  time steps are non-uniform in that the duration of carrier sensing is smaller than the duration of data transmission. In this paper, we introduced a non-uniform time-step formulation of DQN to address this issue. Although we only focus on the use of the modified DQN algorithm for wireless networking, we believe the non-uniform time-step DQN can also find use in other domains, e.g., the Treasury bond investment problem as mentioned in this paper. 

The CS-DLMA framework in this paper assumes the saturated scenario in which all the nodes always have packets to transmit. This will be the case, for example, when the nodes are transmitting large files containing many packets. In other practical scenarios, some nodes may be unsaturated in that they only have packets to transmit intermittently. It will be of interest to investigate CS-DLMA that can deal with heterogeneous networks with a mix of saturated nodes and unsaturated nodes in the future.

\appendices 
\section{}\label{sec:appendix}
This appendix compares the performance of RNN and FNN in CS-DLMA design. In Section \ref{sec:coexist_TDMA_ALOAH}, we show that CS-DLMA with RNN architecture can find the optimal strategies for different $\alpha$  values. In this appendix, we also consider the coexistence of one CS-DLMA node with one TDMA node and one ALOHA node. The settings are the same as in Section \ref{sec:coexist_TDMA_ALOAH} except that we use the FNN architecture instead of RNN in CS-DLMA. In particular, the FNN with two hidden layers is the same as the RNN as introduced in Section \ref{sec:set_up} except that we replace the LSTM layer in the RNN with a feedforward layer. For FNN with more hidden layers (e.g., 10, 20 and 40), we adopt the residual network structure as in \cite{yu2019deep}. The reason to use the residual network structure is to avoid potential overfitting due to large numbers of hidden layers \cite{he2016deep}. 

\begin{figure*}[h]
	\centering
	\includegraphics[scale=0.72]{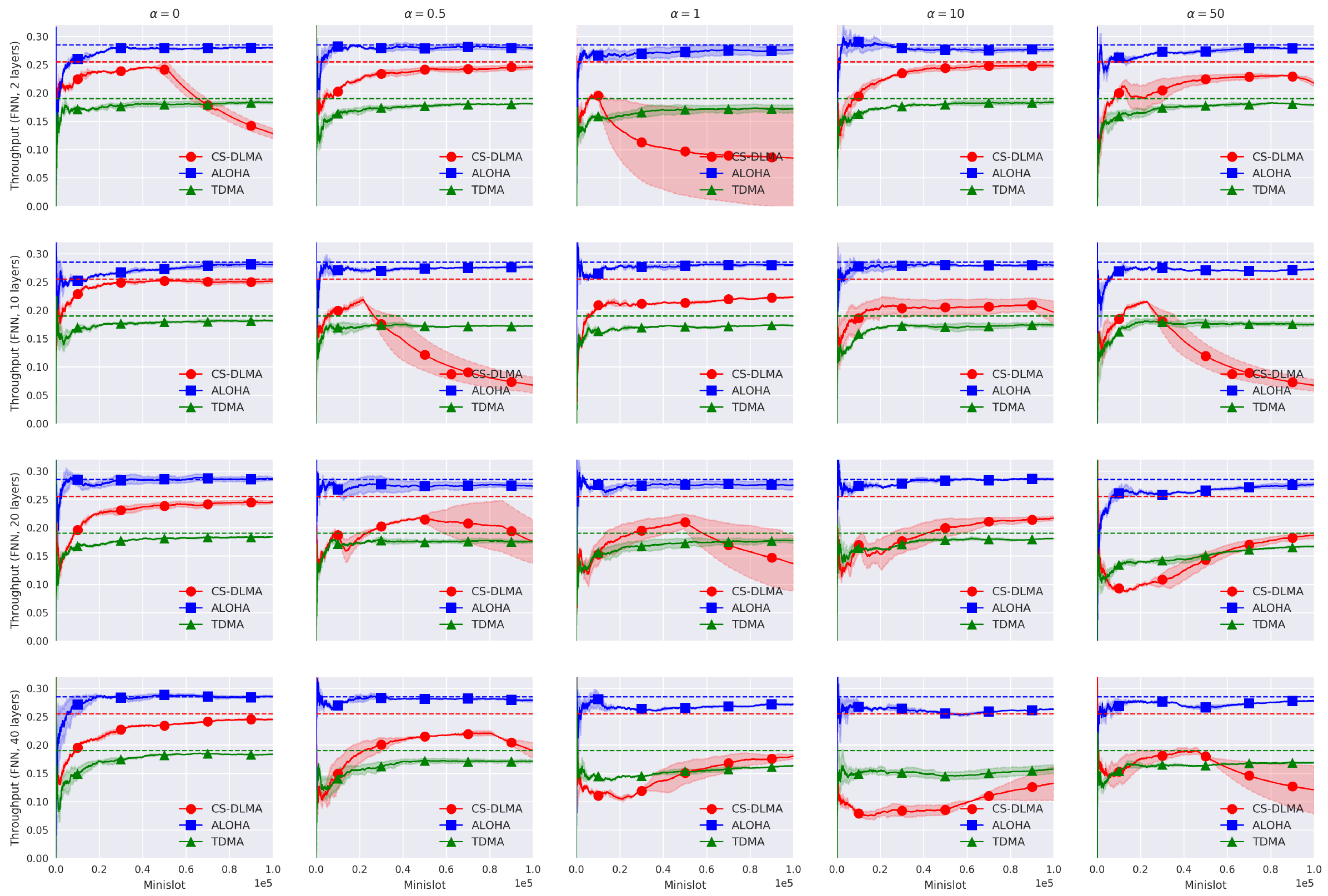}
	\caption[]{Individual throughputs of CS-DLMA, TDMA, and ALOHA for different  $\alpha$ values. CS-DLMA adopts FNN with different numbers of hidden layers. The solid lines are the throughputs achieved by CS-DLMA. Each solid line is averaged over 10 different runs, with the shaded areas being areas within the standard deviation. The dashed lines are for benchmarking purposes. They are the node throughputs when the CS-DLMA node is replaced by a model-aware node operating a model-aware optimal strategy.}
	\label{fig:eva_FNN}
\end{figure*}

Fig. \ref{fig:eva_FNN} presents the individual throughputs of CS-DLMA, TDMA and ALOHA, and their corresponding optimal results. In particular, for different rows in Fig. \ref{fig:eva_FNN}, CS-DLMA uses different number of hidden layers; for different columns, we test the performance of CS-DLMA for different $\alpha$  values. As can be seen from Fig. \ref{fig:eva_FNN}, CS-DLMA with FNN fails to find the optimal strategies for most of the cases, while from Fig. 6 in Section \ref{sec:coexist_TDMA_ALOAH},  we can see that CS-DLMA with RNN can find the optimal strategies for different $\alpha$   values. 

As mentioned earlier in Section \ref{sec:implementation}, the causal relationship between different elements in the input is explicitly modeled into RNN but not FNN. we conjecture that this allows RNN to search within a narrower solution for a good solution (i.e., RNN only needs to learn within a smaller space, allowing it to learn a good solution in a more focused manner). 

\section{}\label{sec:appendix-B}
This appendix derives the benchmark for the case of one CS-DLMA node coexisting with one TDMA node and one ALOHA node---these nodes adopt the settings as introduced in Section 5.2: the CS-DLMA node can transmit packets of variable length with a maximum of 10 minislots; the TDMA node occupies the second and the fifth TDMA slots within a TDMA frame of five TDMA slots; the ALOHA node transmits with a fixed probability $q=0.5$  in each ALOHA slot; and the packet durations of TDMA and ALOHA are both fixed at 10 minislots. 

To derive the benchmark, we imagine a model-aware node that is aware of the MAC details as well the packet durations of TDMA and ALOHA. We replace the CS-DLMA node with this model-aware node in the setting described in the previous paragraph and examine the network performance that can be achieved by this model-aware node. Given that the packet durations of TDMA and ALOHA are the same, we assume that the TDMA slots and the ALOHA slots are aligned in time.  In the rest of this appendix, ``slot'' refers to the TDMA/ALOHA. 

The transmission pattern of TDMA is fixed and not probabilistic. We can divide slots into two categories according to the usage pattern of TDMA: 1) slots occupied by TDMA and 2) slots not occupied by TDMA. For 1), the optimal strategy of the model-aware node is “not to transmit” for any value of  $\alpha$ (transmissions by the model-aware node in these slots will result in collisions and will not contribute to the throughput of TDMA, ALOHA, or the model-aware node).  For 2), we can simplify this problem as the coexistence of the model-aware node with one ALOHA node. 

In general, when coexisting with the ALOHA node, the model-aware node has two strategies---one of which can be the optimal strategy for a particular value of  $\alpha$. These two strategies are given as follows:
\begin{itemize}
	\item \textbf{Greedy strategy:} the model-aware node transmits in all slots of category 2), which results in the throughput of the ALOHA node being zero.  
	\item \textbf{Polite strategy:} the model-aware node first performs carrier sensing in the first minislot and then decides whether to transmit in the next 9 minislots based on the carrier sensing result: if the channel is sensed idle (i.e., ALOHA is not transmitting), then the model-aware node transmits a packet in the next 9 minislots; if the channel is sensed busy (i.e., ALOHA is transmitting a packet in the current slot), then the model-aware node keeps silent in the next 9 minislots.   
\end{itemize}

We can calculate the individual throughputs of the model-aware node and ALOHA node in an ALOHA slot for these two strategies, and the results of are summarized in Table \ref{table:appendix}. 

\begin{table}[htbp]
	\centering
	\renewcommand\arraystretch{1.2}
	\caption{Individual throughputs of the model-aware node and the ALOHA node in an ALOHA slot.}
	\label{table:appendix}
	\begin{tabular}{|l|l|l|}
		\hline
		& Model-aware & ALOHA \\ \hline
		Greedy Strategy & $ 0.475 $ & 0 \\ \hline
		Polite Strategy & 0.425 & 0.475   \\ \hline
	\end{tabular}
\end{table}

It is obvious that the polite strategy is the optimal strategy for any value of  $\alpha$. Therefore, the optimal strategy of the model-aware node for this particular case can be concluded as follows:

From the results shown in Table \ref{table:appendix}, it is obvious to conclude that the polite strategy is the optimal strategy for any value of  $\alpha$. Therefore, the optimal strategy of the model-aware node when coexisting with one TDMA node and one ALOHA node using the settings in Section 5.2 can be concluded as follows:

\textit{At the beginning of each TDMA/ALOHA slot, the model-aware node performs carrier sensing. If the channel is idle, the model-aware node transmits in the next 9 minislots; if the channel is busy, the model-aware node keeps silent in the next 9 minislots. }

Based on the above strategy, the individual throughputs of the model-aware node, the TDMA node, and the ALOHA node can be calculated as 0.255, 0.19, and 0.285, respectively.


%



\ifCLASSOPTIONcaptionsoff
  \newpage
\fi

%

%
%
%


{\footnotesize 
	\bibliographystyle{IEEEtran}
	\bibliography{CS_DLMA}
}

\end{document}